# Possible Atmospheric Diversity of Low Mass Exoplanets – some Central Aspects

John Lee Grenfell[1], Jeremy Leconte[2], François Forget[3], Mareike Godolt[4], Óscar Carrión-González[4], Lena Noack[5], Feng Tian[6], Heike Rauer[1,4,5], Fabrice Gaillard[7], Émeline Bolmont[8], Benjamin Charnay[9] and Martin Turbet[8]

(1) Institut für Planetenforschung (PF)
Deutsches Zentrum für Luft- und Raumfahrt (DLR)
Rutherfordstr. 2
12489 Berlin
Germany

(2) Laboratoire d'Astrophysique de Bordeaux (LAB)
Université Bordeaux
Centre National de la Recherche Scientifique (CNRS) B18N
Alle Geoffroy Saint-Hilaire
33615 Pessac
France

(3) Laboratoire de Météorologie Dynamique (LMD)
Institut Pierre Simon Laplace Université
4 Place Jussieu
75005 Paris
France

(4) Zentrum für Astronomie und Astrophysik (ZAA)
Technische Universität Berlin (TUB)
Hardenbergstr. 26
10623 Berlin
Germany

(5) Institut für Geologische Wissenschaften
Freie Universität Berlin (FUB)
Malteserstr. 74-100
12249 Berlin
Germany

(6) Macau University of Science and Technology
Taipa, Macau

(7) CNRS Orléans Campus
Institut des Sciences de la Terre d'Orléans (ISTO)
1A Rue de la Ferollerie Campus Géosciences
45100 Orléans
France

(8) Observatoire Astronomique de l'Université de Genève
Observatoire de Genève
Chemin de Pégase, 51
1290 Versoix
Switzerland

(9) Observatoire de Paris (OP)
Site de Meudon – 5
Place Jules Janssen
F-92195 Meudon cedex
Paris

*Abstract*: *exoplanetary science continues to excite and surprise with its rich diversity. We discuss here some key aspects potentially influencing the range of exoplanetary terrestrial-type atmospheres which could exist in nature. We are motivated by newly emerging observations, refined approaches to address data degeneracies, improved theories for key processes affecting atmospheric evolution and a new generation of atmospheric models which couple physical processes from the deep interior through to the exosphere and consider the planetary-star system as a whole. Using the Solar System as our guide we first summarize the main processes which sculpt atmospheric evolution then discuss their potential interactions in the context of exoplanetary environments. We summarize key uncertainties and consider a diverse range of atmospheric compositions discussing their potential occurrence in an exoplanetary context.*

1. **Introduction**

Exoplanetary science at the dawn of the 2020s lies at a fascinating juncture at which the basic atmospheric properties of potentially rocky worlds lying in the habitable zone are already beginning to be constrained (see e.g. de Wit et al., 2018 which focused on TRAPPIST-1 planets in the Habitable Zone). Earlier works such as Forget and Leconte (2014) have discussed the potential diversity of exoplanetary atmospheres. Some new aspects were recently summarized by Jontof-Hutter (2019) and Madhusudhan (2019). Tinetti et al., (2018) summarized a chemical survey of exoplanets with the Atmospheric Remote-Sensing Infrared Exoplanet Large-Survey (ARIEL) mission. In the present review we highlight some recent aspects relevant for potential exoplanetary diversity for terrestrial-type exoplanets up to ~10 Earth masses. It is beyond the scope of our work however to review all aspects of this rapidly developing field.

There are numerous developments which motivate our review. First, improved mass-radius data for cooler mini gas planets and super-Earths (see e.g. Fulton et al., 2017; Fulton and Petigura, 2018; Weiss and Marcy, 2014) are confirming that exoplanetary mass can vary over several orders of magnitude for a given radius whereas radius can vary by up to a factor of ~4 for a given mass. Small planets are being found which lie above the pure water composition line in the mass radius diagram - which suggests they have atmospheres. Emerging measurements are providing initial observational constraints for the atmospheres of hot Super Earths. Such new data is providing first hints of atmospheric diversity and is driving emerging strategies for addressing the challenging degeneracies involved as discussed in e.g. Dorn et al. (2017) and Dorn and Heng (2018). Second, progress in detection methods has been made for several atmospheric species. These include theoretical studies involving spectroscopy of collisional pairs as a potential proxy for $O_2$-rich (Schwieterman et al., 2016) and $N_2$–rich atmospheres (Schwieterman et al., 2015; new proposed pathways for abiotically produced CO and $O_2$ (Wang et al., 2016); observational constraints for escaping helium (Spake et al., 2018) on a giant exoplanet as well as new predictions for the mass and characteristic timescales of thin silicate atmospheres on magma ocean worlds (see e.g. Kite et al., 2016). Third, progress has been made regarding key processes such as escape (Owen, 2019) (including early water loss, see Tian et al., 2018) and outgassing (e.g. Gaillard and Scaillet, 2014) over evolutionary timescales. These are being complimented by a growing database of observationally constrained escape rates from exoplanetary atmospheres. Fourth, there have been numerous studies on newly-discovered planetary systems such as TRAPPIST-1 (Gillon et al., 2017) and Proxima Centauri-b (Anglada-Escudé et al., 2016). With the above in mind, we focus in our review on a specific selection of potential exoplanetary atmospheric compositions e.g. based on hydrogen, oxygen, water, silicate, halogen, hydrocarbon, sulfur etc. species and discuss their potential properties and feasibility.

Section 2 introduces key processes and their uncertainties. Section 3 gives a brief overview of atmospheric diversity in the Solar System. Section 4 discusses some key phenomena affecting exoplanetary diversity. Section 5 discusses individual exoplanetary atmospheres. Section 6 presents the conclusions.

**2. Key Processes**

We provide here an initial brief summary of some key processes affecting atmospheric diversity. These include (1) *gas accretion* of growing protoplanets in the disk. Key uncertainties here are the timescales for protoplanetary growth and gas evaporation (Lammer et al., 2018); the role of planetary migration and the dynamical mechanism(s) for volatile delivery (Jacobson and Morbidelli, 2014); (2) Primary (catastrophic) *outgassing* at the end of the magma ocean phase and secondary (volcanic) outgassing. Uncertainties here are related to material properties, convection and composition of the interior; (3) *Delivery* via impacts. Uncertainties involve the effect of impactor properties (mass, composition, radius) over time (De Niem et al., 2012); (4) Atmospheric *escape* which can be broadly split into *thermal* (Jeans, hydrodynamic) and *non-thermal* (sputtering, pickup etc.) escape (see e.g. Tian, 2015a). A key uncertainty here is the evolution of stellar XUV (from 10-91.2nm) which drives atmospheric escape (see e.g. Johnstone et al., 2015; Tian, 2015a). Related processes important for atmospheric development include biogeochemical cycling, weathering, biomass emissions, climate and photochemistry.

**3. Lessons from the Solar System for Exoplanetary Science**

The modern and early Solar System offer valuable lessons for understanding atmospheric diversity. Figure 1 provides a general overview:

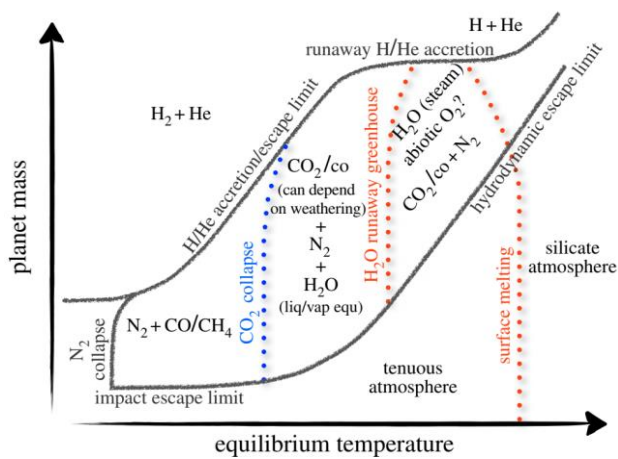

Figure 1: Schematic of atmospheric diversity taken from Forget and Leconte (2014).

We now briefly discuss Solar System atmospheres with reference to the regions in Figure 1.

**3.1 Tenuous atmospheres** – are defined here as objects with surfaces pressure less than a few tens of microbars. They are favored for smaller bodies with weak, gravitational fields or/and strong incoming insolation. Examples are modern Mercury, the Moon, the Galilean Satellites, Pluto and Triton. Table 1 (Grenfell, 2009 and references therein unless otherwise stated) provides a brief overview:

| **Body** | **Species** | **Species Amount** | **Total Surface Pressure** | **Reference** |
|---|---|---|---|---|
| Mercury | Helium | ~$3 \cdot 10^{11}$ molecules cm$^{-2}$ | <$10^6$ molecules cm$^{-3}$ | Grenfell (2009) and references therein |
| | Calcium | ~$1.3 \cdot 10^{11}$ molecules cm$^{-2}$ | | |
| | Sodium | ~$(1-2) \cdot 10^{11}$ molecules cm$^{-2}$ | | |
| | Argon | ~$(0.5-1.2) \cdot 10^{9\$}$ molecules cm$^{-2}$ | | |
| Moon | Helium | ~$(2-40) \cdot 10^3$ molecules cm$^{-3}$ (P$_o$) | $8 \cdot 10^4$ molecules cm$^{-3}$ | Stern et al. (1994) |
| | Argon | $4 \cdot 10^4$ molecules cm$^{-3}$ (P$_o$) | | |
| Io | Sulfur dioxide | $10^{-12}$-$10^{-7\#}$ surface in bar (P$_o$) | | |
| Europa | Atomic oxygen | $10^{-11}$ bar (P$_o$) | | Hall et al. (1995) |
| Ganymede | Molecular oxygen | ~$(1-10) \cdot 10^{14}$ (P$_o$) | <$2 \cdot 10^{-11}$ (bar) | Hall et al. (1998) |
| Callisto | Carbon dioxide | $7.5 \cdot 10^{-12}$ bar (P$_o$) | | Carlson (1999) |
| Pluto | Molecular nitrogen | 0.98 vmr (P$_o$) | ~$1 \cdot 10^{-5}$ (bar) | Grenfell (2009); Wong et al. (2017) |
| | Methane | $10^{12}$-$10^{13}$ molecules cm$^{-2}$ | | |
| Triton | Molecular nitrogen | | $1.9 \cdot 10^{-5}$ (bar) | Gladstone et al. (2016)°° |

Table 1: Overview of tenuous atmospheres in the Solar system. P$_o$ denotes surface value. $^\$$Uncertain, weak outgassing source. $^{\$\$}$Stern et al. (1994), their Table 6. $^\#$Variable due to freeze-out at high-latitudes, Yung and DeMore (1999). °°N$_2$-dominated.

The extremely thin atmospheres ("collisionless exospheres") can be strongly influenced by sputtering from the solar wind (a source of helium e.g. for Mercury and the Moon, Table 1) and by surface mineral properties. The recent *MErcury Surface, Space ENvironment, GEochemistry and Ranging (MESSENGER)* mission revolutionized knowledge of Mercury's alkali metal exosphere (see e.g. Merkel et al., 2017) and near-atmosphere gas-phase ion concentrations (Raines et al., 2014) useful for interpreting atmospheric evolution. Atmospheres of the Galilean moons could be influenced by sputtering onto ice surfaces which splits water and releases oxygen-containing compounds (Table 1).

**3.1.1 Exoplanetary context**

Tenuous atmospheres have been proposed for ultra-short period (USP) (periods of <1day) hot Super-Earths (SEs) such as CoRot-7b (Léger et al., 2009), Kepler-10b (Batalha et al., 2011), 55 Cancri e (Demory et al., 2011) and HD80653 (Frustagli et al., 2020) plus several others (see e.g. Guenther and Kislyakova, 2020 and references therein). These highly-irradiated worlds could represent the stripped cores of smaller hot gas planets but possibly retaining thin atmospheres e.g. derived from refractory material such as silicate (see discussion in Lopez and Rice, 2018) whereas larger, less irradiated SEs could retain their volatile envelopes. Planetary population diagrams (e.g. Fulton et al., 2017; Lundkvist et al., 2016) suggest a radius minimum separating rocky SEs from mini gas planets which is a focus of numerous modeling studies (e.g. Mordasini, 2020). A central issue is whether hot SEs formed mostly in-situ or underwent inward migration which likely depends upon the central properties and timescales of the protoplanetary disk (see e.g. Raymond and Cossou, 2014; Martin and Livio, 2016). Direct observations of the proposed tenuous atmospheres of hot SEs is rather lacking although proxy data for atmospheric mass and composition is emerging.

Regarding Corot-7b earlier model studies (e.g. Schaefer and Fegley, 2009) suggested silicate atmospheres with strong sodium and calcium absorption lines analogous to Mercury and Io. Kurokawa and Kaltenegger (2013) suggested that XUV-driven photoevaporative loss could remove one Jupiter mass of gaseous envelope on Corot-7b (and Kepler 10b) within ~1Gyr. Guenther et al. (2011) placed upper limits for alkali metals based on atmospheric spectral features for CoRot-7b. Samuel et al. (2014) implied that obtaining a few transits with the James Webb Space Telescope (JWST) could extend such analyses by constraining atmospheric mass and composition of CoRot-7b. Regarding Kepler-10b, Rouan et al. (2011) suggested the observed phase curves were best fitted by a rather high bond albedo of ~0.5 with a tenuous ($p_o$<2mb) silicate atmosphere (see also Herbort et al. (2020) who discuss theoretical atmospheres which can form above a (partially) melted crust). By comparison Esteves et al., (2015) derived mostly lower albedos of <0.25 a sample of 14 close-orbiting Kepler mission planets

Regarding 55 Cancri e, phase curve data (Tsiaris et al., 2016) based on the Hubble Space Telescope (HST) Wide Field Camera (WFC3) suggested a hydrogen-helium dominated atmosphere with the surface pressure posterior peaking at 100mb (although this quantity was only loosely constrained) and a possible detection of hydrogen cyanide (HCN). High resolution transit spectroscopy (Ridden-Harper et al., 2016) suggested a modest (3$\sigma$) sodium detection which would be consistent with an exosphere extending to ~five planetary radii. Spitzer thermal flux observations (Demory et al., 2016[b]) suggested a strong (~1300K) gradient between the dayside and the nightside of 55 Cancri e which implies either a thin atmosphere or weak transport from day to night. Assuming partial vaporization and surface exchange from molten magma pools, Kite et al. (2016) suggested a tenuous atmosphere with a surface pressure of ~10Pa (~$10^{-5}$ bar) on 55 Cancri e. Angelo and Hu (2017) however suggested a rather thick atmosphere with a surface pressure of ~1.4 bar based on a model fitted to time series of Spitzer IR photometry. Hammond and Pierrehumbert (2017) applied a General Circulation model to 55 Cancri e which suggested that a light (90:10, $H_2$:$N_2$) atmosphere with several bars surface pressure could best fit the phase curve data. Lightcurve observations including revised stellar activity (Bourrier et al., 2018) however did not support such a lightweight atmosphere. Miguel (2019) and Zilinskas et al. (2020) subsequently modeled the detectability of atmospheric spectral features by JWST assuming heavy, nitrogen-dominated atmospheres on 55 Cancri e varying [C/O] from 0.01 to 200. Results suggested e.g. that HCN should be clearly detectable for high C/O scenarios. Variability in 55 Cancri e lightcurves (e.g. Demory et al., 2016[a]; Tambura et al., 2018 and Sulis et al., 2019) could arise due to e.g. stellar variability or due to a possible dust torus. The latter work suggested upper constraints on the optical albedo (<0.47 at 2-sigma) for 55 Cancri e based on the non-detection of secondary eclipses in visible light with the Microvariability and Oscillation of Stars (MOST) satellite. Folsom et al. (2020) modelled the stellar wind based on Zeeman Doppler Imaging and concluded that 55 Cancri e orbits inside the Alfvén surface which suggests that planet-star interactions could occur. Dorn et al. (2019) applied an interior model using a Bayesian approach which suggested that 55 Cancri e could have a mantle enriched in calcium and aluminium and may even lack a planetary core which would have poorly-known consequences for atmospheric evolution via outgassing. Modirrousta-Galian et al. (2020) suggested strong tidal forces could trap light atmospheres on the nightside where they could survive stellar driven escape.

In summary, a range of atmospheric masses and compositions have been suggested based on observations of Cancri 55e over the last decade. This could arise because such measurements approach the capability limits of modern instrumentation, or possibly due to intrinsic variability in the atmosphere via e.g. volcanism, or due to external variability via e.g. stellar luminosity or dust. Follow-up observations with JWST should help address these issues.

Tenuous exo-atmospheres are favoured targets in the sense that they could be sampled *over their entire vertical extent* by exoplanetary transmission spectroscopy. For thicker atmospheres one typically samples only the upper regions since the underlying layers (pressures in the range of about (10 -100) millibars depending on wavelength and composition) become optically-thick or the light is refracted away from the observer (García Muñoz et al., 2012; Bétrémieux and Kaltenegger, 2014). Tenuous atmospheres will therefore more likely offer atmospheric windows sampling down to the surface potentially revealing surface mineralogy (although identifying spectral signals of exoplanetary surfaces is observationally very challenging, see Hu et al., 2012; Madden and Kaltenegger, 2020) which via sputtering could influence the main atmospheric constituent - as on Mercury (Table 1). Also, such worlds could have extensive tails of e.g. sodium extending out to several planetary radii - as on Mercury (Potter and Killen 2008) and on the Moon (Matta et al., 2009). Such "cometary-like tails" have also been found for Hot Neptunes such as GJ436b in the form of escaping hydrogen (Kulow et al., 2014; Ehrenreich et al., 2015) and for super-Mercuries in the form of dust (Budaj, 2013). Upper atmosphere detections of atomic species (although currently most such detections have so far been achieved mostly for hot Jupiters, see Madhusudhan et al., 2016 and references therein) could be early probes of atmospheric composition.

Close-orbiting planets are on the one hand favored targets due to short orbital periods which suggest a high number of transit events over a given period. On the other hand, tenuous exo-atmospheres are generally favored by smaller bodies with smaller transit depths which vary proportional to the factor ($r_p^2/r_*^2$). Also, low mass atmospheres in general could lead to weak planetary absorption features. Korablev et al. (this issue, chapter 14) and Tinetti et al. (this issue, chapter 16) discuss future observations of exoplanetary atmospheres from space.

**3.1.2 Lessons for Exoplanetary Diversity –** for Mercury-like worlds, observations from the Solar System suggest atmospheres could consist of (1) a helium component depending on e.g. planetary protection and the impinging stellar wind,(2) a gas-phase alkali metal component depending on e.g. the availability of alkali minerals and surface mineralogy and (3) a smaller (about x100 less) amount of outgassed noble gases. Regarding point (1), Vidotto and Bourrier (2017) (their Table 1) summarize estimates of stellar mass loss rates and atmospheric extent for different stellar classes. Boro Sakia et al. (2020) suggest that current model estimates of stellar wind properties such as mass and momentum loss rates could vary by a factor (2-10). Regarding point (2), the theoretical study by Hu et al. (2012) suggested that JWST could possibly identify e.g. surface silicate IR bands for some nearby targets although this will be very challenging.

**3.2 Silicate Atmospheres**

The inner rocky planets and the Moon experienced an early period of intense bombardment lasting for a few hundred million years after formation. This led to local and global magma ocean (MO) events. Zahnle et al. (2010) suggested early hot rock (silicate) atmospheres having characteristic removal timescales of up to ~1000 years with silicate rainout proceeding via the condensation rock sequence (e.g. corundum, forsterite, wüstite etc.). The order of condensation is uncertain depending upon the reducing nature of the MO-atmosphere system which is likely sensitive to the rate of iron rainout into the core as it formed. Pahlevan et al. (2011) discussed hot rock rainout on early Earth calculated with a two-phase thermodynamical model having magnesium rich droplets in an iron rich vapor. Fegley and Schaefer (2012) reviewed studies of Earth's silicate vapor atmosphere including rainout, (silicate) snow formation and in-situ photochemical reactions. Somewhat lacking in the literature are calculations of silicate atmosphere rainout using modern, time-dependent microphysical models. Direct spectral observations of silicate atmospheres on e.g. hot rocky SEs are very challenging but could be achieved with JWST (see also discussion in section 5.1.)

**3.3 Steam Atmospheres**

Atmospheres dominated by water vapor represent a critical phase for the early development of terrestrial planets. The timing and interplay between key processes such as escape and outgassing during the steam atmosphere phase can strongly influence the onset and duration of the planet's subsequent habitability. The nature and magnitude of outgassing at the end of the MO phase is related to the progressive exclusion of volatiles from the minerals driven by bottom-up solidification and depends on cooling rates and convection turnover timescales (see e.g. Maurice et al., 2017; Boukaré et al., 2018). Crustal formation occurred on Earth between ($10^4$-$10^7$) years after the final global MO phase began (see e.g. Lebrun et al., 2012; Nikolaou et al., 2019). The range of uncertainty arises mainly due to poorly-constrained material properties of the mantle and uncertainties in atmosphere-interior coupling.

Regarding Venus-Earth-Mars (VEM) Lammer et al. (2018) and references therein (their Table 2) suggested catastrophic outgassing of (450, 200, 50) bars of water on VEM respectively. Lebrun et al. (2013) estimated that the first global oceans condensed (~0.1, 1.5, 10) Myr thereafter on VEM respectively. Way et al. (2016) suggested early Venus could be habitable up to t=715 Myr assuming one tenth of an Earth ocean based on their 3D model study. Lammer et al. (2018) however suggested that on Venus the steam atmosphere may never have condensed and the water could have been possibly lost via early escape. On Earth, numerous studies (e.g. Elkins Tanton et al., 2008; Elkins-Tanton et al., 2012; Hamano et al., 2013; Lebrun et al., 2013; Massol et al., 2016; Salvador et al., 2017) investigated how the early thick steam atmosphere formed, cooled, condensed and ultimately formed the oceans. Model studies (e.g. Katyal et al., 2019) investigated cooling of Early Earth's steam atmosphere due to outgoing longwave (LW) radiation at different stages of the MO.. Earth proxy data (ancient zircons) support early oceans 4.3 to 4.4 Gyr ago (Mojzsis et al., 2001). Direct spectral observations of giant steam atmospheres in an exoplanet context are currently lacking.

**3.4 Carbon Dioxide ($CO_2$) - Nitrogen ($N_2$) - Oxygen ($O_2$) Atmospheres**

After the onset of Earth's oceanic and continental formation, carbon dioxide was increasingly subject to atmospheric removal via washout and weathering. The role of atmospheric escape weakened as the early stellar EUV output decreased (e.g. Ribas et al., 2005). Other long-term processes also began to play a role for subsequent atmospheric evolution. These involved wide-ranging and sometimes subtle feedbacks within the interior-hydrosphere-lithosphere-biosphere-atmosphere system (see e.g. Spohn, 2014). Processes such as secondary outgassing, weathering (continent formation), climate, photochemistry, feedback cycles and eventually biology and biogeochemical cycles of carbon, nitrogen and oxygen (discussed below) start to influence atmospheric evolution. Charnay et al., (this issue, chapter 3) review climate evolution on the Early Earth and the faint young sun problem.

**3.4.1 Volatile Inventory of Earth's chondritic building blocks**

To gain insight into the volatile budget of the basic building blocks of our planet, Tables 2 and 3 show the gas compositions at p=100 bar which are in equilibrium with the metal chondrites (Table 2) and the carbonaceous chondrites (Table 3) at 500K, 1000K and 2000K. Data is from Zahnle et al. (2010) their Figures 2 and 3 which are based on Schaefer and Fegley (2010):

| Species | %Molar 500K | %Molar 1000K | %Molar 2000K |
|---|---|---|---|
| $NH_3$ | 1.2 | <1% | <1% |
| $N_2$ | <1% | <1% | <1% |
| $H_2S$ | <1% | <1% | 1.3 |
| $CO_2$ | <1% | <1% | 2.4 |
| $H_2O$ | <1% | 6.3 | 18.6 |
| CO | <1% | <1% | 25.7 |
| $H_2$ | <1% | 25.1 | 47.9 |
| $CH_4$ | 94.4 | 64.6 | <1% |

Table 2: Gas compositions (%molar) at 100 bar which are in equilibrium with the metal chondrites (Schaefer and Fegley, 2010).

| Species | %Molar 500K | %Molar 1000K | %Molar 2000K |
|---|---|---|---|
| $NH_3$ | <1% | <1% | <1% |
| $N_2$ | <1% | <1% | <1% |
| $H_2S$ | <1% | <1% | 4.2 |
| $CO_2$ | 66.1 | 21.9 | 15.1 |
| $H_2O$ | 28.2 | 66.1 | 63.1 |
| CO | <1% | 1.4 | <1% |
| $H_2$ | <1% | 6.0 | 6.6 |
| $CH_4$ | 5.9 | <1% | <1% |
| $SO_2$ | <1% | <1% | 2.0 |

Table 3: As for Table 2 but for the carbonaceous chondrites.

Table 2 (iron chondrites) suggests the chemical equilibrium favors volatile speciation into methane and ammonia at lower temperatures, shifting to hydrogen, carbon monoxide and water at higher temperatures. Table 3 (carbonaceous chondrites) suggests that chemical equilibrium favors mainly carbon dioxide with some water at lower temperatures, shifting to mainly water with some carbon dioxide together with sulfur compounds and hydrogen at higher temperatures. Fegley and Schaefer (2012) discuss theories for constraining the relative amounts of chondritic material making up the bulk Earth.

**3.4.2 Venus Earth Mars**

Numerous texts have discussed the present-day (e.g. Yung and DeMore, 1999) and evolutionary compositional development (e.g. Lammer et al., 2018) of $CO_2$-$N_2$-$O_2$ atmospheres on Venus, Earth and Mars (see also Lammer et al., 2020). Regarding Venus e.g. Lammer et al. (2018); Hamano et al., (2013) and Lebrun et al., (2013) suggested that strong insolation compared to the Earth may have lengthened the steam atmosphere phase for long enough (~100 Myr) such that efficient hydrodynamic escape could have effectively dried out the planet. In this scenario, the thick, hot, modern $CO_2$ atmosphere on modern Venus results directly from outgassing at the end of the MO and *not* via the more frequently discussed runaway moist greenhouse effect. Regarding $N_2$, nitrogen isotope data (Marty, 2012) favors a carbonaceous chondrite origin. Some nitrogen in the form of $NH_3$ (Lammer et al., 2018 and references therein) could have existed near the Venusian surface especially if the early atmosphere was shielded from UV.

Regarding Earth, weaker insolation than Venus and long-term plate tectonics maintained surface oceans with stabilizing climate feedbacks involving atmospheric $CO_2$ such as the carbonate-silicate cycle (Walker et al., 1981) (Charnay et al., this issue) discuss climate on Early Earth and the faint young Sun problem). The build-up of $N_2$-$O_2$ dominated atmospheres are favored by the presence of life (Lammer et al., 2018).

Regarding Mars, escape mainly drives atmospheric evolution. Odert et al. (2018) assumed outgassing of ($CO_2$=11bar) and ($H_2O$=85bar) at the end of the MO period and then calculated complete atmospheric loss via escape after 18 Myr (25 Myr) assuming a moderately (slowly) rotating early Sun. Supporting this, xenon data from ancient Martian meteorites suggest that early Mars lost most of its initial atmosphere (Cassata, 2017). Observations of modern Mars by the Mars Atmosphere and Volatile Evolution (MAVEN) spacecraft (Jakosky et al., 2018) suggested integrated loss rates (assuming Solar evolution) on Mars of 0.8 bar $CO_2$ and a 23m water ocean layer. Kite et al. (2019) review constraints in the evolution of atmospheric pressure and liquid water on early Mars based on geological data. Warren et al. (2019) suggested a continuous surface pressure upper limit of 1.9 bar at 4 Gyr on Mars based on an analysis of small craters. High Martian $^{15}N$ compared to the other rocky planets suggests that Mars lost a significant amount of its early nitrogen inventory (Lammer et al., 2018). Regarding outgassing, the reduced Martian interior could favor enhanced outgassing of e.g. $H_2$ and CO compared with Earth (Ramirez et al., 2014). This could lead to an interior-atmosphere negative feedback whereby outgassing of reduced species on Mars leads to enhanced atmospheric loss of the reductant H which drives oxidation of the atmosphere and hence (via e.g. deposition) ultimately favors mantle oxidation over time. Catling et al. (2003) discussed similar mechanisms operating on Early Erath.

**3.4.3 Isotopic fractionation**

Several of the key process mentioned above e.g. escape, delivery and interior processes, result in isotopic fractionation. Numerous advances leading to reduced data precision for isotopic fractionation ratios have occurred in recent years. For example, Hin et al. (2017) suggested that accretional vaporization played an important role in the volatile budget of terrestrial planet formation based on improved isotopic magnesium ratios. New ruthenium isotope data (Fischer-Gödde and Kleine, 2017) suggested surprisingly that the late veneer featured an *inner* solar system origin and was likely not the primary source of volatiles on Earth. Improved noble gas and nitrogen isotope data has advanced knowledge of early escape, especially on Mars (Cassata, 2017; Füri and Marty, 2015) whereby observed neon and argon isotope ratios (Odert et al., 2018) support a slow to moderately rotating (rather than a quickly rotating) early Sun.

**3.4.4 Exoplanetary Context**

Inspecting volatile inventories of terrestrial planet building blocks (Tables 2 and 3) based on chondrites from the Solar System gives a general indication of the potential range of atmospheric species in an exoplanetary context where such observations are not available. Hints on exoplanetary mineral compositions can be gained from e.g. stellar metallicity measurements coupled with interior structure plus atmospheric escape modeling (Dorn and Heng, 2018). Planet formation and interior structure models (Alibert and Venturini, 2019) also provide some insight. Initial observations of infalling material from circumstellar debris onto white dwarfs as reviewed in Farihi (2016) have been performed. Results suggest exomineral compositions which are not dissimilar to those of the Solar System. Adibekyan et al. (2015) investigated observed [Mg/Si] ratios of [F,G,K] stars using HARPS. Their results suggested an evolution in this ratio through the Galaxy and that smaller mass stars have higher ratios. We now discuss briefly the key processes affecting atmospheric diversity of terrestrial exoplanets.

Regarding *escape*, progress is somewhat hindered by a lack of various parameters such as stellar EUV observations, stellar wind properties and planetary exospheric temperatures which can all influence atmospheric escape. Improved estimates of such parameters e.g. improved stellar EUV by semi-empirical modeling (Fontenla et al., 2016) is useful in this regard. Section 4.3 discusses (extreme) escape processes in more detail. Regarding *outgassing*, numerous factors could be relevant which we will briefly discuss in approximate order of observing difficulty. The bulk planetary density can be estimated already today for some SEs by measuring planetary radius from the transit depth together with planetary mass from the radial velocity method. The planet's mass distribution throughout the interior can be in principle constrained by determining the love number although observations are presently limited to large, close-in exoplanets (e.g. Hellard et al., 2020). The core mantle ratio and interior composition can be constrained from Bayesian analyses (Dorn et al., 2015) and machine-learning (Baumeister et al., 2020) techniques driven by observations of e.g. planetary mass, radius, love number and stellar metallicity. Knowledge of surface composition e.g. via reflection spectroscopy could provide hints on mantle convection. For example active plate tectonics on Earth produces continental crust mainly in the form of granite (see discussion in Herbort et al., 2020, their section 3.2). Hu et al (2012) analyzed detectability of reflection and thermal emission spectra on "airless Earths" for a range of mineral surfaces including basaltic, ultramafic and granitoid. Distinguishing the resulting small signals (they suggested a few parts per million differences in the ratio of planetary to stellar flux for e.g. Kepler 20f) however is likely beyond the reach of JWST and is more an issue for spectroscopy missions further into the future. Other factors which could influence outgassing include the amount of water in the interior which could affect mantle convection. The water content of Earth's mantle is not well constrained, ranging from (~0.2 to 3.0) Earth ocean masses (Ohtani, 2020) and is not known for rocky exoplanets. Recent works (e.g. Namouni and Morais, 2020) suggest that high inclination Centaurs in our Solar System could have an interstellar origin so could represent the first opportunity to constrain the building blocks of exoplanets, although such a task is very challenging.

Another interesting effect impacting outgassing is the change in the mantle oxidation state (oxygen fugacity) over time (see e.g. Gaillard et al, this issue; Kadoya et al., 2020; Gaillard et al, 2015) which can in turn depend on global processes which change the Earth's redox balance and influence outgassing. Observational constraints of mantle composition (hence fugacity) in an exoplanet context are however weak as stated above. Over time, disproportionation reactions ($3Fe^{2+} = Fe° + 2Fe^{3+}$) in the interior associated to the oxygen-friendly nature of minerals growing in the transition zone and lower mantle (Frost and Mc Cammon, 2008) that may induce an oxidation of the upper mantle. This would however be conditioned by the efficient removal of the produced metallic iron (Fe°) into the forming core and by the efficiency of global convection in mixing lower and upper mantle materials. These mass transfer processes are in turn conditioned by the establishment of plate tectonics representing an efficient convection regime at planetary scale. In brief, the physical status of the planet interior (style of convection, mantle size) and the chemical state, which rules the efficiency of mantle outgassing, are intimately intermingled. This implies that planetary mass and outgassing could be somehow related. Broader relationships exist between planetary interiors and surficial processes. For example, in the upper atmosphere, loss to space of atomic hydrogen from e.g. water photolysis escapes to space, must leave behind oxygen-enriched atmosphere, leading to oxidizing continental and seafloor weathering, which if introduced in the mantle by subduction processes could influence its redox state and outgassing efficiency.

Can we test the above feedbacks - proposed by theory and occurring between e.g. atmospheric escape and the planetary interior - in an exoplanetary context with observables? Regarding escape, an evaporating tail can form in the trailing planetary limb which could influence the transit lightcurve duration and shape e.g. in the Lyman-alpha for escaping H-atoms. Oza et al. (2019) summarize spectroscopic detections in the exospheres of hot Jupiters. Chen et al. (2020) discuss detection of sodium, magnesium and H in the hot exosphere of WASP-52b and review species detections in the exoplanetary exospheres. Ehrenreich (2015) estimated from transit observations escape rates of ($10^8$ to $10^8$) g/s for the warm Neptune GJ436b. For the potentially rocky SEs in the TRAPPIST-1 system, Bourrier et al., (2017) estimated H-escape rates of (4.6, 1.4, 0.6) $x10^7$g/s for TRAPPIST-b,c,e respectively based on Lyman-alpha transit data obtained with HST. Whether the presence of a magnetic field increases or decreases atmospheric escape rates is debated (see Dehant et al. 2019, their section 4.2). Exoplanet science could soon actively contribute to this debate. Proposed techniques involving radio emission and polarimetry to constrain exoplanetary magnetospheres are under discussion (e.g. Zarka et al., 2019). Proxy approaches have provided first constraints e.g. magnetohydrodynamic simulations needed to reproduce observed excursions in peak brightness to the east and the west of the substellar point on the hot Jupiter HAT-P-7b, required a planetary magnetic field of at least 6 Gauss (Rogers, 2017) (compared with ~0.5 Gauss on Earth's surface). Cauley et al., (2019) estimated strong magnetic fields in the range (20-120) Gauss for four HJs based on measurements of the Calcium II K line. The potential role of escape in atmospheric evolution is discussed in section 4.3 below.

How central interior processes such as mantle convection, plate tectonics, outgassing etc. respond to changing planetary mass such as for SEs is still a subject of debate (see e.g. Noack et al., 2017; Stamenkovic and Breuer, 2014; Fratanduono et al., 2018), but this brief description shows that the style of geo(exo)dynamics is pivotal in ruling the nature of outgassed atmosphere. Godolt et al. (2019) modeled rocky planets in the HZ of early M-dwarf stars which have very long active pre-main sequence phases. Their work suggested that planetary *regassing* could be important for re-instating habitability after the long (~1Gyr) intense (~x10-1000 EUV of the modern Sun).

Regarding delivery of *water* and other volatiles in exoplanetary systems, some key issues are e.g. identifying the main processes affecting planetary formation and orbital migration (see discussion in Morbidelli and Raymond, 2016); the extent of turbulent mixing in the disk (Furuya et al., (2013); determining the position of the water, ammonia and carbon dioxide ice lines (see e.g. Pinilla et al., 2017). The role of dynamical perturbations [e.g. the "Grand Tack" (Jacobson and Morbidelli, 2014)] which may be important for delivering volatiles to terrestrial planets which form dry, is not well known for different exoplanetary systems as a function of stellar and exoplanetary properties of e.g. exo-Jupiters and exo-Saturns. Note that these objects are expected to be discovered by the ongoing Gaia mission. Regarding mini gas planets, water-dominated atmospheres are favored thermodynamically (T>~500K, P>~a few bar) for solar metallicities of ~x1000 (Moses et al., 2013, their Figure 4). More model formation studies are required investigating the effect of impacts upon atmospheric delivery and escape as a function of disk and planetary parameters. Direct observations of impact-driven delivery in exoplanetary atmospheres are not available. Next generation missions however observing dust distributions in protoplanetary and debris disks, as proposed for the Wide-Field Infrared Survey Telescope (WFIRST) (Nuecker et al., 2016; Douglas et al., 2018) could however help constrain bombardment rates of early exoplanets by asteroids which could ultimately help improve the planetary formation models.

Regarding *$CO_2$ atmospheres*, the presence of an Earth-like ocean with active plate tectonics and a hydrological cycle could enhance $CO_2$ removal via washout, formation of carbonate and subduction. For weak weathering cases without plate tectonics, e.g. Foley and Smye (2018) suggested possible runaway $CO_2$ scenarios. Responses of $CO_2$ on rocky exoplanets are generally challenging to predict, being subject to potentially complex process in the carbon cycle which are mostly unconstrained in an exoplanetary context. Climate stabilizing cycles are sensitive to many factors including the land-sea mask (e.g. Lewis et al., 2018) and planetary orbital parameters (Williams and Pollard, 2002). Regarding exoplanets $CO_2$ has been detected for several planetary atmospheres (mainly HJs) and JWST will aim to expand knowledge of this species in terrestrial-type atmospheres (Greene et al., 2016). Obtaining observational hints for oceans (e.g. Robinson et al., 2014) and surface minerals (see e.g. Hu et al., 2012; Cui et al., 2018, as discussed) are however challenging. Regarding mini gas planets, $CO_2$ atmospheres are favored thermodynamically (T>~500K, P>~a few bar) for solar metallicities of ~x10,000 (Moses et al., 2013, their Figure 4).

Regarding *$N_2$ atmospheres*, global sources and sinks are not well known (see e.g. Lammer et al., 2019; Wordsworth, 2016). Numerous processes such as volatile delivery (Hutsemékers et al., 2009), outgassing, lightning, cosmic rays and biological activity all play a role. Further exoplanetary studies are required to establish which key processes affect atmospheric $N_2$ evolution (see also section 5.7). Direct spectral observations of $N_2$ in exoplanetary atmospheres are currently lacking since its spectral features are weak. Schwieterman et al. (2015) discussed constraining nitrogen abundances via spectral absorption from $N_2$-$N_2$ collisional pairs. Oklopčić et al. (2016) discussed identifying $N_2$ and $H_2$ via so–called ghost line features in Raman spectroscopy. These signals however correspond to changes of up to a few percent in planetary geometric albedo in the UV so are beyond the reach of present day missions.

Regarding *$O_2$ atmospheres*, e.g. Gumsley et al. (2017) discuss relevant processes on Early Earth such as changes in carbon burial rates, mantle redox, biological activity etc. which could influence atmospheric oxygen abundance. Lingam and Loeb (2019) review the potential for photosynthesis on Earth-like planets. Abiotic $O_2$ production involving e.g. water photolysis followed by hydrogen atom escape (Luger and Barnes; 2015; Tian, 2015b; Wordsworth et al., 2018) or via $CO_2$ photolysis followed by self-reaction of oxygen atom (Gao et al. 2015, Tian et al., 2014) is an expanding field. Note however that NOx and HOx species (as on Mars) and additionally ClOx (as on Venus) can act as efficient catalysts to facilitate the formation of $CO_2$ from CO and O as discussed above. In this case, the $N_2$, $O_2$, $H_2O$ and UV amounts in the planetary atmosphere (required to form NOx and HOx) could become important. Grenfell et al. (2018) suggested that atmospheric $O_2$ abundances can be upper-limited by explosion-combustion. Theoretical studies (e.g. López-Morales et al., 2019; Rodler and López-Morales, 2014) suggested that several tens of transits are needed with the ELT to detect atmospheric $O_2$ assuming a close-by transiting Earth-like atmosphere. Assuming an Earth-like atmosphere on our (non-transiting) neighbor Proxima Centauri-b would require about 55 hours (assuming R=150,000) ELT observing time (Hawker and Parry, 2019) to detect the atmospheric $O_2$ A-band in reflected light. Fauchez et al., (2020) suggested that atmospheric $O_2$ could be detected via a relatively strong collision-induced absorption feature in the mid/far-infrared (6.4 microns) and may thus be accessible with JWST.

**3.5 Hydrogen ($H_2$) and Helium (He) atmospheres**

ProtoEarth and ProtoVenus grew to (50-75%) (Lammer et al., 2018 and references therein) of their present planetary mass by ~10 Myr and accreted thereby (0.1-10%) $M_{earth}$ hydrogen envelopes i.e. up to a factor of ~$10^5$ times present Earth's atmospheric mass, after which time the gas in the protoplanetary disk had mostly evaporated. Earth's primordial light atmosphere was likely formed mainly by impact-induced volatile degassing but also by direct gravitational accretion of gas from the disk (Abe, 2011). Escape processes driven by uncertain, early XUV evolution from the star are likely important for deciding timescales and to which extent the primordial atmosphere is lost (e.g. Kislyakova et al., 2015).

### 3.5.1 Jupiter, Saturn, Uranus and Neptune

Table 4 shows the composition by volume of the dominant atmospheric species ($H_2$ and He) for the giant planets in the Solar system compared with the Solar Values:

| Species | Solar | Jupiter | Saturn | Uranus[*] | Neptune[*] |
|---|---|---|---|---|---|
| Hydrogen ($H_2$) | 0.912 | 0.897 | 0.963 | 0.85 | 0.80 |
| Helium (He) | 0.085 | 0.102 | 0.033** | 0.15 | 0.19 |

Table 4: Fractional composition by volume for the Sun (Grevesse et al., 2007), Jupiter and Saturn (Bagenal et al., 2004) and Uranus and Neptune (Schmude, 2008). [*]Middle atmosphere measurements.

**Jupiter and Saturn** - small differences between Jovian and Solar composition in Table 4 are suggested to arise due to helium differentiation in Jupiter's core. First-order phase transitions from molecular to metallic hydrogen could also affect bulk composition (see e.g. Guillot, 1999). The Cassini mission considerably advanced knowledge of dynamical and temperature responses on Jupiter (e.g. Flasar et al., 2004) and Saturn (Dyudina et al., 2008). Helium amounts determined from this mission (Koskinen and Guerlet, 2018) suggest more abundant values, $vmr_{He}$=0.09 in the lower atmosphere compared with the earlier data shown in Table 4. Atmospheres of Jupiter and Saturn feature complex hydrocarbon chemistry (e.g. Smith and Nash, 2006) and cloud formation via e.g. phosphorous, water, and sulphur-containing species (Atreya et al., 1999).

**Uranus and Neptune** – likely have small rocky cores and are generally carbon-enriched compared with Jupiter and Saturn (Ali-Dib et al., 2014; Mousis et al., 2018). Recent advancements have been made in understanding the circulation (Orton et al., 2014) and brightness (Nettelmann et al., 2015) of Uranus as well as the composition (e.g. Moreno et al., 2017) of Neptune. Hydrocarbon photochemical responses on the ice giants are discussed in Moses et al., (2019).

Atreya et al., (2020) comprehensively reviewed planetary composition, structure, origins and potential missions for the Solar System Gas and Ice Giants.

### 3.5.2 Exoplanetary context

When interpreting the bulk composition of the Solar System's giant planets, it is interesting to separate the influence of formation timescales and migration on the one hand, from that of planetary parameters such as mass, radius and core properties on the other hand. Studying cool exoplanetary giant planets in future will be useful in this regard. Helium detections in exo-Neptunes such as HAT-P-11b (e.g. Mansfield et al., 2018; Allart et al., 2018) are relevant in this respect. Hu et al. (2015) modeled helium exo-atmospheres on warm Neptunes such as GJ436b. Regarding atmospheric hydrocarbons, model studies (Moses et al., 2018) suggest summer-winter asymmetry due to photochemical responses in the atmospheres of the giant planets in the Solar System which they suggested could be tested in an exoplanetary context using the JWST if suitable cool giant exoplanet targets are found. Regarding (Ultra) Hot Jupiters [(U)HJs], their atmospheric composition and detection are reviewed in e.g. Maldonado et al. (2018) and Madhusudhan (2019); their atmospheric structure in Laughlin (2018) and transport in e.g. Komacek and Showman (2016). Grenfell et al. (2020) summarized potential atmospheric observations of HJs and nearby SEs using the planned color filters on the PLATO fast cameras. We do not however discuss (U)HJs in detail here since our focus in the present work lies with the smaller mass terrestrial-type exoplanets

**3.6 Summary** – the Solar System may be atypical in some respects. For example it features neither mini gas planets nor SEs which may be rather common in the Universe (Burke et al., 2015). Nevertheless, studying the Solar System provides valuable lessons regarding e.g. volatile budgets, delivery, magma oceans, outgassing and steam atmospheres for the inner terrestrial planets as well as lessons regarding the formation and composition of outer gas planets. This can be an invaluable exercise when considering the potential diversity of exoplanetary atmospheres.

**4. Key Phenomena anticipated to affect Exoplanetary Atmospheric Diversity**

**4.1 Elemental abundances in the protoplanetary disk**

Molecular clouds (MCs) are one of the most investigated phenomena in the interstellar medium, yet the main mechanisms by which they form, evolve and collapse are not well understood (see overview by Vázquez-Semadeni et al., 2006). MCs typically contain ~90% $H_2$ gas with masses in the range ~$(10^2-10^6)M_*$ and temperatures (10-50K). Differential rotational rates in the branches of the Milky Way can lead to MC rotation. At the rotational equator of the MC, the outward centrifugal force balances the inward gravitational force. Along the MC rotational axis however, the unbalanced gravitational force leads to inward transport of material so that the MC flattens into an accretion disk (Maeder, 2009). Observations of evolving disks by the Atacama Large Millimeter Array (ALMA) telescope have greatly advanced the field (Matrà et al., 2018).

Metallicity distribution functions across the Milky Way were discussed e.g. in Nidever et al. (2014) and Hayden et al. (2015). The latter study suggested (Fe/H) values ranging from -0.5 to 1.5 for a sample of ~70,000 stars. Variations in elemental abundances across the protoplanetary disk, however, depend on uncertainties in the dynamics and timescales of e.g. disk collapse, disk mass, dust-to gas ratios and photochemical effects. Predicting the position of disk snowlines and the resulting variation of planetary elemental ratios such as (C/O) and (C/H) in the gas and solid phases helps constrain whether such species were accreted as dust or gas during planetary formation. This can in turn help differentiate between proposed planetary formation mechanisms such as the 'gravitational instability' as opposed to the 'core accretion' mechanism (see Venturini and Helled, 2017). It can also provide information on migration rates and constrain location(s) in the disk at which the forming planet accreted its metals (see e.g. Booth et al., 2017; Krivov and Booth, 2018; Lin et al., 2018; Booth and Ilee, 2019). Dorn et al. (2015) used stellar elemental data and planetary mass-radius observations to constrain core-mantle radii and discussed future applications of this technique to constrain exoplanetary atmospheres and oceans. Several theoretical studies have varied metallicity and modeled the resulting atmospheres assuming chemical equilibrium (see review by Madhusudhan et al., 2016 and references therein). Moses et al., (2013) calculated the effect of varying metallicity values from x1 to $x10^4$ Solar in hot Neptunes and calculated (their Figure 5) atmospheres ranging from $H_2$ and $H_2O$ for medium-range (x1-$x10^2$ Solar) metallicities and forming $N_2$ atmospheres for higher metallicities. For metallicities greater than x100 Solar, they noted $CH_4$-dominated atmospheres for temperatures less than 600K and CO-dominated atmospheres for temperatures greater than 800K. Hu and Seager (2014) varied assumed mole fractions, $X_H$ (from 0 to 1) and $(X_C/X_O)$ (from 0 to 1 to 10) in a model study of the SE GJ1214b. Their results suggested atmospheres ranging from e.g. highly oxidized (e.g. $O_2$-dominated) at low (C/O) changing to highly reduced ($CH_4$ and other hydrocarbons such as $C_2H_2$ (ethyne) and $C_2H_4$ (ethene) at high (C/O)).

**4.2 Variations in initial interior volatile reservoirs and outgassing**

Processes which influence outgassing can be broadly considered as follows. First, the initial (local) volatile concentration delivered to the planet, which is related to the concept of feeding zones (Grewal et al., 2019) and the changes in orbital trajectory during planetary accretion (see 3.4.4). Then, the C-H-O-N-S elements which are supposedly outgassed from the planetary interior to form the atmosphere are not always volatiles. Recent research efforts (Hirschmann 2012; Rohrbach et al., 2014; Dasgupta and Grewal, 2019; Malavergne et al., 2019) clearly show that these elements can be siderophile or refractory, implying that under certain conditions or during specific stages in the differentiation of a planet, they are not outgassed from the mantle, but remain sequestrated in the planetary interior. Technically, the effective partitioning of volatiles into the melt depends on e.g. redox state, melt fraction, composition, temperature and pressure. These parameters are linked to the exogeodynamics (see 3.4.4) defining the style of planetary internal convection (eg. Noack et al., 2014). On planet Earth, geochemical constraints indicate that carbon is in the oxidized form of carbonate in the shallow mantle (ie. <180 km depth) while it is completely in the form of diamond at greater depth (Frost and Mc Cammon, 2008; Gaillard et al., 2015). This redox transition is critical for the removal of carbon upon mantle melting since only oxidized carbon can be outgassed while diamond/graphite is refractory. This also has implications on the speciation of other volatiles like hydrogen, by affecting reduced volatiles such as $H_2$ and $CH_4$ (e.g. Dasgupta, 2013). On Mars, oxidized carbon is not expected due to a much reduced mantle; this implies a limited carbon outgassing from the Martian mantle in comparison to Earth. Finally, the volatility of the C-H-O-N-S (transfer from melt to gas) is the last condition for outgassing into the atmosphere. Nitrogen and sulfur are not volatile under the most reduced conditions typical of Mercury for example (Gaillard et al., 2015). Under oxidizing conditions, however, they can be as volatile as other components (ie. $CO_2$, CO, $H_2O$, $H_2O$). Outgassing $CO_2$ and CO from the melt is always favored since these are very volatile species. Water is much less volatile than $CO_2$. This hierarchy in the volatility implies that the composition of the outgassed mixture from a basalt is chiefly pressure dependent: at high pressure (~a few tens of bar and above), water is not outgassed, while at low pressure (< 0.1 bar), most of the C-H-O-N-S elements can be outgassed (Gaillard and Scaillet, 2014). In brief, pressure and oxygen fugacity are the parameters ruling the amount and nature of degassed mixture from a basalt.

How could mantle oxygen fugacity have evolved through time on planet Earth? Frost and McCammon (2008) suggested an oxidized mantle was reached at ~1Gyr after core formation. Li and Lee (2004) and Trail et al. (2011) provided evidence for constant $fO_2$ values in the upper mantle over geological time back to 4.3 Ga. A consensus seems to emerge here, but there remain large uncertainties preventing the used redox poxies to provide firm conclusion (e.g. Yang et al., 2014). During core-mantle separation, conditions were extremely reduced implying an atmosphere dominated by $H_2$ and CO (Hirschmann, 2012; Gaillard et al., 2015). After core formation, the efficiency of carbon removal into the core influenced the subsequent mantle $fO_2$ and coeval atmosphere evolution (Hirschmann, 2012). How oxygen fugacity evolved during the late stage of the magma ocean appears pivotal (Elkins-Tanton et al., 2008; Elkins-Tanton et al., 2012). Recent efforts link the depth of the magma ocean to the evolution of its oxidation state and the fate of carbon (Armstrong et al., 2019). It is then possible that right at the end of the magma ocean stage, the oxygen fugacity of the Earth's mantle was similar to the present-day one with an oxidized shallow mantle and a deep reduced one. But such planetary systems may not always be considered as a closed system: Over geological timescales, (a) prolonged hydrogen atmospheric escape (involving water photolysis followed by removal of hydrogen to leave behind oxygen which can be weathered into the surface), (b) delivery of oxidized late veneer material and (c) subduction of oxidized mineral could also drive mantle oxidation (Kasting et al., 1993). Orders of magnitude difference in the outgassed mass of volatiles can arise depending on the processes listed above.

**4.3 Extreme atmospheric escape**

Atmospheric escape can be limited by diffusion (depending on e.g. atmospheric scale height of different species and the concentration of escaping species) or energy (depending on incoming XUV or high energy particle fluxes from the central star). Regarding atmospheric escape in the modern Solar System, Jupiter is sufficiently massive and far from the Sun to prevent efficient escape in virtually any form. Regarding Venus, Earth and Mars, escape processes have been reviewed by numerous works e.g. Shizgal and Arkos (1996). In the early Solar System, atmospheric escape on Venus, Earth and Mars was discussed in e.g. Lammer et al. (2018). On Pluto, Zhu et al. (2014) suggested energy-limited escape which hydrodynamic treatments under-estimate by ~13% (see also Hoey et al., 2017). The New Horizon observations suggest that theoretical calculations of atmospheric escape on Pluto generally over-estimate observed losses.

Regarding exoplanets, numerous earlier works modeled hydrodynamic escape on Hot Jupiters (see Tian, 2015 and references therein). Tripathi et al. (2015) applied a 3D hydrodynamic model which suggested day-night asymmetries in outflow for tidally-locked Hot Jupiters. Weber et al. (2017) modeled plasma conditions near the exobase of Hot Jupiters. Insight into planetary formation and the role of escape has been revealed by observational studies of the so-called Neptunian desert (e.g. Mazeh et al., 2016) and the "radius" or "Fulton" gap (Fulton et al., 2017; Owen and Wu, 2017; Lehmer and Catling, 2017; Jin and Mordasini 2018; Gupta and Schlichting, 2019) for close-in planets. Lammer et al. (2013) suggested blowoff and Roche lobe overflow could occur for SE atmospheres. Tian (2009, 2013) modeled planetary upper atmospheres containing heavy molecules such as $N_2$, $O_2$ and $CO_2$ under strong XUV radiation. Johnstone et al., (2015) and Erkaev et al., (2016) modeled the important role played by XUV for escape processes for SEs with hydrogen dominated atmospheres.

Regarding rocky exoplanets orbiting M-dwarf stars, an important question is to what extent they can retain their atmospheres. Tian and Ida (2015) suggested that Earth-like planets with Earth-like oceans could be rare for cool star systems due to efficient water loss over the extended pre-main sequence. The model study by Owen and Mohanty (2016) included an updated escape model with improved radiative cooling and transition between hydrodynamic and ballistic regime, which suggested that large primordial hydrogen envelopes may survive the early, extreme early EUV phase for rocky worlds around cool stars. Dong et al. (2018) modelled the TRAPPIST-1 system and suggested ion escape rates ($10^2$-$10^3$) times faster than that on Earth assuming an initial one bar atmosphere with Venus-composition. Their modelled stellar winds were ~three times faster than the Sun for the same star-planet distance. Their work suggested complete atmospheric removal for the inner TRAPPIST-1 planets for surface pressures up to ~10bar. They noted that more work is required to characterize the properties of the incoming stellar wind. It should also be noted that the impact of rapid loss on the energy budget of the underlying planetary upper atmosphere is not considered in Dong et al. (2018) and thus their loss rate could be more limited by the energy conservation principle proposed in Tian (2013). Kubyshkina et al., (2018) presented an updated method based on gridded analytical fits to replace the energy-limited formula commonly employed to estimate atmospheric escape. Tian et al. (2018) reviewed recent progress in understanding the evolution of stellar XUV and the relevance for water loss on terrestrial exoplanets. Johnstone et al. (2018) developed a flexible 1D upper atmosphere escape model validated for Earth and Venus to model the effect of e.g. stellar XUV evolution. Applying this model to terrestrial-type planets orbiting early, active solar-like stars (Johnstone et al., 2019) suggested Earth-like atmospheres cannot form due to transonic hydrodynamic escape of e.g. C and O.

**4.4 Clouds**

Clouds are common features in planetary atmospheres throughout the Solar System (see e.g. Taylor, 2010). Cloud microphysical processes include homogeneous nucleation, heterogeneous nucleation (occurring on condensation nuclei such as aerosol and dust particles), coagulation (growth) and sedimentation. Complex climate feedbacks exist between clouds, radiation and transport which are still being unraveled in Earth's atmosphere (see e.g. Ceppi and Hartmann, 2015).

In exoplanet science, the central role of clouds and hazes is becoming more and more apparent. The presence of clouds can flatten the transmission spectra of Hot Jupiters (see e.g. Parmentier et al., 2016; Wakeford et al., 2016; Powell et al., 2018). Regarding hot SEs, Mahapatra et al., (2017) investigated the potential role of atmospheric mineral-based clouds. Schaeffer and Fegley, (2009) suggested clouds consisting of sodium or/and potassium-containing minerals for hot SEs which (for favourable candidates) could be observable with current facilities.

Regarding warm SEs, Kreidberg et al. (2014) suggested the presence of clouds in the atmosphere of GJ1214b in order to be consistent with transit spectroscopy data (see also Charnay et al., 2015; Mollière et al., 2017; Gao and Benneke (2018). Morley et al. (2015) modelled cloudy thermal emissions and reflection spectra in SE atmospheres. Kawashima et al. (2019) performed feasibility studies for spectral detections by the James Webb Space Telescope (JWST) assuming clouds and hazes on GJ1214b, GJ 436b, HD 97658b, and Kepler-51b. Benneke et al. (2019) suggested that Mie-scattering by clouds is responsible for the flat spectrum they observed for GJ3470b at visible and near infrared wavelengths. The impact of such clouds should decrease for observations at longer wavelengths.

**5. Case Studies of Exoplanetary Atmospheres**

**5.1. Magma Ocean Worlds**

A magma ocean (MO) planet features a molten mantle without a planetary crust and surface temperatures in the range ~(1000-2000K) which is generally sufficient to melt silicate minerals depending on their composition. Such conditions can be reached temporarily e.g. after the planetary accretional phase, or permanently e.g. for close-orbiting terrestrial exoplanets as discussed in Hamano et al. (2013) and Nikolaou et al. (2019). A number of potential MO exoplanets have been identified, such as Corot-7b, 55 Cancri e, TRAPPIST-1b and Kepler 10b (e.g. Hammond and Pierrehumbert, 2017; Henning et al., 2018; Nikolaou et al., 2019 and references therein).

Exoplanetary model studies of MO atmospheres focused on e.g. interior-atmosphere coupling, atmospheric composition, cooling processes and the length of the MO phase (see e.g. Lebrun et al., 2013). Kite et al. (2016) suggested MO world atmospheres close to vapor-pressure equilibrium determined by fractional vaporization and surface-interior exchange. Marcq et al. (2017) applied a coupled interior-atmosphere model simulating $H_2O$-$CO_2$ atmospheres which confirmed that the outgoing LW blanketing effect saturates at the Nakajima limit (Nakajima et al., 1992) of ~280 $Wm^{-2}$ (and ~40% less for atmospheres assuming clouds) for scenarios with 300 (100) bar surface pressure of $H_2O$ ($CO_2$). They suggested the limit breaks down for surface temperatures above 1690K (1970K) for scenarios with (without) clouds, although these values depend on atmospheric composition and mass. Breakdown occurs as successively hotter planetary surfaces emit radiation at successively lower wavelengths. Lupu et al. (2014) modeled atmospheric composition and evolution of post giant impact atmospheres on Earth-like planets. Nikolaou et al., (2019) applied a coupled interior-atmosphere model to discuss whether a sample of potential MO exoplanets could lie in temporary ("evolutional MO") or permanent MO states depending on e.g. albedo and incoming stellar radiation.

Atmospheric species on MO worlds can be removed directly by surface ingassing e.g. for noble gases as indicated by studies of mantle $^3He/^{22}Ne$ preserves (Tucker and Mukhopadhyay, 2014). Sharp (2017) suggested ingassing during the MO phase could represent a major mechanism for delivery of volatiles into the interior (see also Olson and Sharp, 2018). Wu et al. (2018) discussed mantle acquisition of noble gases, $H_2$ and $H_2O$ during the MO phase. Wordsworth (2016) discussed thermolytic dissociation or/and dissolution of nitrogen-containing species into the MO. Kite et al (2019) suggest that ingassing of $H_2$ during the MO phase may be responsible for forming the steep decrease in the planetary occurrence rate between 3-4 Earth radii.

**5.2 Steam Atmospheres**

A massive steam atmosphere e.g. with hundreds to thousands of bars pressure at the surface can be outgassed during crustal formation at the end of the MO phase on terrestrial-type planets. Subsequent cooling could then lead to ocean formation via condensation. Elkins-Tanton (2011) noted that planetary formation models suggest SEs could commonly acquire their oceans via such a mechanism. On the other hand, if the planet's incoming stellar radiation is strong, this could result in a prolonged steam atmosphere phase with atmospheric desiccation via water photolysis followed by escape. Alternative to steam outgassing, it is feasible that so-called waterworlds (Léger et al., 2004) could evaporate their water inventory and form massive steam atmospheres if they migrate sufficiently far inwards. Kite and Ford (2018) suggested that waterworlds could be common and modeled their habitability limits. Goldblatt (2015) modeled the dependence of steam atmosphere amount and duration upon planetary mass and incoming stellar radiation. Fegley et al. (2016) suggested that exoplanetary steam atmospheres could dissolve magnesium, iron and alkali metal oxide minerals which could potentially modify their surface composition. The model study of Pluriel et al. (2019) suggested high bond albedos (~0.8) for steam exo-atmospheres ($T_{surf}$~1000K) (an effect associated with clouds) and medium bond albedos (~0.5) for hotter steam atmospheres (($T_{surf}$~2000K) where gas absorption becomes more important.

Steam atmospheres are also investigated in the context of determining via model studies the inner habitable zone boundaries (HZ). Earlier works e.g. Kopparapu et al. (2013) noted the importance of employing detailed, up-to-date line-lists for water when determining the inner HZ. Leconte et al., (2013) investigated simulating in 3D the effect of under-saturated descending Hadley cells. A summary of 1D and 3D modeling studies of steam atmospheres near the inner HZ, including important cloud feedbacks are summarized in Godolt et al., (2016) and references therein. Thomas and Madhusudhan (2016) noted that thick steam atmospheres could increase the radii of water-rich planets by up to several tens of percent - which could be used as an observational indicator for the volatile content on sampled rocky exoplanets by future missions (see also Turbet et al., 2019 and Turbet et al., 2020).

**5.3 Oxygen Atmospheres**

The global oxygen cycle (on modern Earth) is influenced by processes extending over the interior-lithosphere-biosphere-atmosphere system (see e.g. Petsch, 2003; Catling and Claire, 2005).

Regarding *sources*, the main net source of atmospheric $O_2$ is via subductive burial (~12 Tmol C/yr; Holland, 2006) occurring mainly near continental margins (Hartnett et al., 1998). This process regulates how much $O_2$ from photosynthesis (forming $O_2$) can enter the atmosphere by subducting organic material, making it unavailable for respiration (destroying $O_2$). A minor source of $O_2$ ($3\times10^8$ molecules cm$^{-2}$ s$^{-1}$) on modern Earth, occurring ~100 times slower than burial; see Yung and DeMore, 1999) is via atmospheric escape which proceed via photolysis of water followed by escape of H then self-reaction of the resulting O to form $O_2$. This $O_2$ source could however be important in Early Earth type environments and on terrestrial-type exoplanets (see below).

Regarding *sinks*, the main sinks of atmospheric $O_2$ (on modern Earth) include (a) atmospheric reaction with reduced outgassed species such as $CH_4$ and $H_2$ (resulting in a net sink of ~1-3Tmol/yr $O_2$; Catling and Claire, 2005), (b) metamorphic reactions of $O_2$ directly on hot rock surfaces (~2.5 Tmol/yr $O_2$; Catling and Claire, 2005) associated with volcanic activity, and (c) weathering (~16 Tmol/yr $O_2$, Holland, 2002) occurring via $O_2$ dissolving in rain followed by chemical reactions on surface rock. Earlier, mainly box model studies have investigated $O_2$ evolution over geological timescales on Earth (see Lenton and Watson, 2000; Berner, 2001). Lacking however in the literature are evolutionary studies with recent, coupled climate-photochemical models. Gebauer et al. (2017) and Gebauer et al. (2018) investigated processes affecting atmospheric oxygen on Earth and on Earth-like exoplanets respectively. On Earth, $O_2$ levels have remained stable at ~21% for several hundred Myr. There is no firm understanding of the feedback mechanisms which maintain long-term regulation of $O_2$ (Lasaga and Ohmoto, 2002).

Regarding exoplanets, lessons from Earth science suggest that processes affecting $O_2$ are numerous and diverse. Regarding exoplanetary $O_2$ *sources*, (a) the development of photosynthesis was likely an evolutionary singularity in the history of our planet, as reviewed by Fischer et al., (2016). Gale and Wandel (2017) (building on earlier work of Kiang et al., 2007) investigated the potential of Earth-like worlds orbiting M-dwarf stars to support exo-photosynthesis. Meadows et al., (2018) discussed interpreting $O_2$ as an atmospheric biosignature in the context of its environment. Lingam and Loeb (2019) discussed the impact of exo-photosynthesis and its possible variants upon the atmospheres of Earth-like planets. Ward et al. (2019) review factors affecting atmospheric oxygen in the Solar System and beyond. The above studies assume that Earth-like planets in the HZ of M-dwarf stars can retain atmospheres although some works e.g. Airapetian et al. (2017) suggested that ion escape could efficiently remove Earth-like atmospheres on such worlds within tens to hundreds of Myr. Regarding (b) the $O_2$ source due to H-escape discussed above, clearly this process is favored for smaller-mass exoplanets orbiting stars active in XUV. Abiotic gas-phase reactions could also influence atmospheric oxygen in Earth-like planets.

Although not significant in Earth's modern atmosphere, abiotic production of $O_2$ could proceed efficiently in e.g. $CO_2$-dominated atmospheres via $CO_2$ photolysis whereby the resulting O-atoms self-react to form molecular oxygen. The abiotic production is strongly disfavored by the presence of hydrogen oxides (HOx) and nitrogen oxides (NOx). These species are produced in Earth-like atmospheres e.g. from lightning and cosmic rays which release HOx and NOx by breaking down their reservoir species such as nitric acid and water. HOx and NOx disfavor abiotic $O_2$ production by driving catalytic gas-phase cycles which convert atomic O (from $CO_2$ photolysis) back into $CO_2$. In an atmosphere with a high concentration of $O_2$ from $CO_2$ photolysis, the $O_2$ is likely to coexist with a high concentration of CO, therefore simultaneous observation of CO and $O_2$ could therefore be an efficient method to distinguish the abiotically-produced from the biological $O_2$ (Wang et al. 2016). The mechanism was discussed in Selsis et al. (2002); Tian et al., (2014); Gao et al., (2015) and Luger and Barnes (2015).

Regarding exoplanetary $O_2$ *sinks*, these likely depend upon (a) outgassing in both amount and composition of reduced compounds. These depend in turn upon mantle dynamics and the ability of the interior to absorb compounds e.g. during core formation, as discussed above. The metamorphic sink (b) will be favored on exoplanets with extended hot rocky surfaces. The weathering sink (c) will be favored on exoplanets with strong hydrological cycles and extensive, long-term continental coverage which are currently unknown parameters. Timescales for establishing the continents calculated by models for early Earth vary from several hundred Myr (fast growth) to several thousand (slow growth) Myr (Taylor and McLennan, 1985) depending on uncertainties in the dynamics and heat distribution in the mantle. Grenfell et al. (2018) suggested an upper limit sink for $O_2$ on Earth-like planets (depending on T,p, composition) due to explosion-combustion initiated by cosmic rays or lightning. Wordsworth et al. (2018) performed modeling studies of GJ1132b, an Earth-sized planet transiting an M-dwarf star. Results suggested $O_2$ atmospheres could vary over a large range - from tenuous atmospheres up to very thick (several thousands of bar at the surface) depending mainly on the initial planetary water budget assumed. Earth-like atmospheres, defined here as ($N_2$-$O_2$) dominated were suggested to be geobiosgnatures by e.g. Lammer et al., (2019), Airepatian et al., 2017) since their build-up together to significant values (>tenths of bar surface pressure or more) likely requires life to generate the required redox disequilibrium. Krissansen-Totton et al. (2016) suggested that strong redox disequilibrium arose on Earth due to the simultaneous presence of $N_2$ together with $O_2$ and liquid water (see also Krissansen-Totton et al., 2018).

In summary, the extent and diversity of $O_2$-dominated atmospheres on Earth-like planets is currently difficult to estimate. Key uncertainties include the likelihood of photosynthesis occurring beyond the Earth and upon abiotic processes occurring in-situ in the atmosphere. Whereas biological $O_2$ formation likely proceeds at the surface, abiotic $O_2$ production likely peaks in the atmosphere depending on the trade-off between its chemical precursors and the availability of UV/XUV. This difference could lead to differing vertical distributions of $O_2$ which if measurable could be a means to distinguish the two mechanisms. Strong and protracted outgassing of reduced compounds together with strong rainout and continental coverage and the presence of plate tectonics could lead to a reduction in the atmospheric $O_2$ abundance via weathering. We now apply the above ideas in order to discuss a current research question.

**5.3.1 What is the fate of oxygen or/and water during runaway climate processes?**

As discussed earlier, for planets that are close to their host stars, a runaway greenhouse could occur and the planet could develop a dense stream atmosphere with a hot climate for a certain period of time. If there is adequate UV radiation to dissociate water molecules and the XUV radiation imposed on the planet is strong enough, rapid loss of H and O will occur. Note that the loss of H has been suggested (Tian, 2015a) reach the ratio of 2:1 so that net loss of water occurs. If the initial loss of H is more than O, atmospheric $O_2$ could build up and lead to more loss of O. If the inventory of water on the planet is massive, it is possible that the planet will remain a water-rich body with a dense steam atmosphere and an extended H and O envelope for an extended period of time. Such types of planets do not exist in our Solar System but might be observed in the future in exoplanetary systems. Observationally these types of planets would be similar to mini-Neptunes except that they have more oxygen in their extended atmospheres. We note that planets with an extended H and O envelope may also be experiencing a moist greenhouse state in which the atmosphere is not dense and steamy.

The initiation and maintenance of planetary (climate) runaway require an imbalance in climate feedbacks in which net positive (self-enforcing) feedbacks (e.g. ocean evaporation-greenhouse heating; ice-albedo) outbalance net negative stabilizing feedbacks (carbonate-silicate cycle; cloud reflection; Planck emission). Some key processes which could prevent a runaway planet from accumulating $O_2$ are as follows. Firstly, $O_2$ may be absorbed into the magma ocean if the mantle is sufficiently reducing (see e.g. Wordsworth et al., 2018). Secondly, the surface could remove $O_2$ via weathering (see 5.3) depending on e.g. mineral composition, or/and the presence of plate tectonics. Processes one or/and two could have occurred on early Venus and Venus-like worlds (see Wong et al., 2019). Thirdly, $O_2$ could rapidly escape in the hydrodynamical regime (see e.g.

Bolmont et al., 2017; Johnstone et al., 2019) e.g. if the star is highly active in XUV, as discussed. Modeling studies are required to disentangle the fate of oxygen and water due to runaway climate effects.

### 5.4 Methane atmospheres

**5.4.1 What would Titan look like closer to the sun?**

We discuss here a thought-experiment in which modern Titan is shifted towards the Sun and then consider how the atmosphere would respond. We consider thereby Titan as an example of an icy body (waterworld) which has migrated inwards from beyond the snowline (see e.g. Léger et al., 2004). The potential habitability of waterworlds was discussed in e.g. Goldblatt (2015) and Kite and Ford (2018). A central aim of our thought experiment is to instigate further discussion and to encourage additional modeling studies of atmospheric evolution on these worlds to understand atmospheric diversity. Note that Turbet et al., (2018) considered how Titan would evolve if placed at the orbit of the TRAPPIST-1 planets. They concluded that such a warm Titan would likely be efficiently depleted in $CH_4$ due to strong photolysis. This is favored since $CH_4$, unlike $H_2O$, continuously exists in "moist greenhouse" since firstly, it has no atmospheric cold trap being hard to condense, and secondly, it absorbs efficiently in the near infrared which weakens the cold trap.

Titan has a cold (94K) surface temperature with ~1.5 bar $N_2$-dominated atmosphere containing ~1.4% $CH_4$ in the stratosphere, up to ~5% $CH_4$ near the surface (Tobie et al., 2005) and thick organic hazes in the middle and upper atmosphere (see e.g. Yung and DeMore, 1999 and references therein). Much was learnt about Titan's atmosphere during the Cassini mission (see e.g. Coustenis et al., 2009). Krasnopolsky (2019) reviews observational and modeling insight. At the start of our thought experiment, instellation begins to increase so that methane in the upper layers is photodissociated e.g. at Lyman-alpha wavelengths to form (in the case of $CH_4$) a mixture of radicals ($C^1D$, $CH$, $CH_2$, $CH_3$) and hydrogen ($H_2$, $H$) species (see Gans et al., 2011) and their ions. Nitrogen is also dissociated in the EUV into nitrogen atoms and ions. Titan's mass is only 6.7% that of the Earth (~3.3 Earth moons) so that the atomic hydrogen resulting from methane photolysis could be efficiently lost as Titan moves inwards. This mechanism could be especially important for early exo-Titans which migrate inwards when the young central star is active in XUV. With increasing temperatures, evaporation of the lighter (short chain) organic aerosol is generally favored over heavier (long chain) aerosols which condense more easily than their lighter chain counterparts. The climate response of the thick haze is however potentially complex since it could either counteract (amplify) warming from increased instellation by reflecting (scattering) incoming shortwave radiation depending on its composition, shape, size distribution and altitude.

An interesting level of instellation is reached in our thought experiment where the surface temperature exceeds $T_{evap(CH4)}$ ~112K at which point liquid methane begins to boil. Lidar data from Cassini (Mastrogiuseppe et al., 2014) suggests Titan's methane-ethane lakes would lead to ~70m mean coverage on Titan's surface but contain only ~1.4% of atmospheric methane. Evaporation would therefore lead to only a modest influx of methane into the atmosphere. This would likely block incoming radiation, strengthen the anti-greenhouse effect and favor haze formation, but increase methane photolysis in the upper layers and hence likely increasing escape rates of hydrogen (a product of methane photolysis). Another interesting level of instellation is reached where water ice or/and ammonia-water layers melt (see e.g. Tobie et al., 2005) which have an estimated mean thickness from 50 to 200km based on Cassini orbital data (see e.g. Hemingway et al., 2013 and references therein). This global melting event could lead to the formation of deep surface oceans containing ammonium hydroxide which has a strongly suppressed freezing point compared to water.

Further increasing instellation could eventually lead to evaporation of the $NH_3$ and $H_2O$ ocean (see Yang et al., 2017). Depending on the UV environment, the $H_2O$ vapor could photolyse to form the hydroxyl (OH) radical which is a strong in-situ sink for atmospheric $CH_4$, whereas the $NH_3$ (if not efficiently shielded from UV) would by photolytically removed (see. e.g. Wolf and Toon, 2010). A plausible scenario for the final atmospheric state is therefore mainly $N_2$ with some water vapor and $CH_4$ depending on the uncertain initial water inventory and rate of atmospheric escape. Such a habitable state could however be potentially short-lived because of the relatively short timescale of H loss.

Although numerous studies have modeled Titan's climate and composition, also with General Circulation Models (see e.g. Lora et al., 2015; Lebonnois et al., 2012) there are much fewer studies investigating Titan's atmospheric response to changing insolation. Lorenz and Lunine (1997) modelled Titan's future atmosphere response as the Sun evolves into a red giant. They proposed an initial insensitivity to increasing solar energy due to the thick organic haze and discussed UV-haze feedbacks as the Sun reddened and weakened in UV output. They also discussed the potential presence of water-ammonia oceans. McKay et al., (1991) discussed potential climate responses on modern Titan due to the net greenhouse heating effect (+15K) which they decomposed into (a) an anti-greenhouse cooling (-9K) due to hazes plus (b) greenhouse heating (+21K) due to collision-induced absorption of $N_2$-$N_2$, $CH_4$-$N_2$ and $H_2$-$N_2$.

In an exoplanetary context, since modern Titan's atmosphere originates either directly from the Saturnian high density disk or/and from cometary impacts early in its history (e.g. Lunine et al., 1998), one can therefore anticipate exo-Titan atmospheres to consist of $N_2$-$NH_3$-$CH_4$-$H_2$ gas mixtures depending on e.g. the location and timescale of formation and the role of cometary delivery. Migration likely depends on numerous factors such as the timing and location of the object's early growth. The presence of a large primordial $H_2$ envelope could strongly influence the photochemistry since $H_2$ participates in "cracking" cycles which shorten the atmospheric hydrocarbon chain, as occurs also on Jupiter (see Yung and DeMore, 1999). On modern Titan, $H_2$ is scarce, so the "cracking" effect is weak and photochemical cycles involving radical-radical combination lead to build-up of long chain hydrocarbons which fall out of the atmosphere and form tar like substances on the surface.

Gilliam and McKay (2011) simulated an exo-Titan with the same instellation as modern Titan but placed around an M4 dwarf star. Results suggested that enhanced incoming IR from the star could penetrate the exo-atmosphere more efficiently since the haze is more transparent at such wavelengths. This led in their study to ~10K surface warming to ~104K which is only ~7K below the boiling point of $CH_4$. One can therefore imagine exo-Titans which are moderately warmer than modern Titan so that the methane lakes have evaporated. Finally, note that proposed cryo-volcanism driven by tidal-heating on modern Titan (see e.g. Tobie et al., 2006) which could supply $CH_4$ into Titan's atmosphere from clathrate hydrates beneath the surface - would of course stop operating had Titan migrated inwards early in its history. This could shorten methane recycling timescales between surface and atmosphere by removing a potentially significant source.

In summary our Titan-based thought experiment suggests that migrated bodies from beyond the ice line could feature oceans containing ammonia-water mixtures with depressed freezing points or/and atmospheres consisting of steam-ammonia-methane-nitrogen mixtures depending on e.g. the origin and rate of migration and the incoming instellation.

**5.5 Carbon monoxide atmospheres**

Carbon monoxide (CO) can be formed in planetary atmospheres by several mechanisms including (a) *in-situ* gas-phase processes e.g. via photolysis of $CO_2$ or (b) outgassing if redox conditions in the mantle are suitable or (c) *thermodynamically* at high pressures and temperatures. For example, in the middle to lower atmosphere of Hot Jupiters and sub-Neptunes, CO can be the thermodynamically favored form of carbon for suitable metallicities.

**5.5.1 In-situ gas phase formation via $CO_2$ photolysis**

$CO_2$ features temperature-dependent photolysis in the UV absorption band at 120<λ<210nm to form $CO(^1\Sigma^+)$ and (mostly) the ground-state $O(^3P)$ triplet together with vibrationally-excited $O(^1D)$ singlet (e.g. Schmidt et al., 2013). The product $CO(^1\Sigma^+)$ can itself photolyse below 167nm. The atomic oxygen triplet can undergo three-body (termolecular) self-combination to form $O_2$. For Earth-like atmospheres orbiting (F,G,K) stars, Selsis (2002) suggested that $O_2$ formed abiotically in $CO_2$-dominated atmospheres could limit its own production by absorbing the same photons needed to split $CO_2$. For planets orbiting M-dwarf stars, however, the extended pre-main sequence and strong early XUV was proposed to drive strong abiotic oxygen production through either (a) photolysis of $CO_2$ followed by self-reaction of oxygen or (b) photolysis of $H_2O$ followed by rapid loss of H (see e.g. Luger and Barnes, 2015) as discussed above. The reverse reaction to the first step in (a) i.e. involving three-body (termolecular) recombination of CO and O is electronically quantum spin forbidden (Jasper and Dawes, 2013) and typically proceeds with a gas-phase rate ~$10^4$ times slower than the forward photolysis reaction in planetary atmospheres (Krasnopolsky, 1982).

Families of gas-phase reactive species involving e.g. hydrogen oxides (HOx), nitrogen oxides (NOx) and chlorine oxides (ClOx) can catalytically drive the recombination of CO and O into $CO_2$. It is convenient to group fast-reacting species into "families" in this way since although the individual species can quickly inter-react, the sum of their concentrations (e.g. HOx=OH+$HO_2$) is removed more slowly from the atmosphere and is therefore conserved (on timescales of ~weeks). Calculating the sum HOx, NOx, ClOx etc. is therefore useful since it indicates the atmosphere's potential to drive catalytic cycles which e.g. form $CO_2$ from its precursors (on Mars and Venus) or to destroy ozone (on Earth). The family members are formed from so-called reservoir species which generally have longer lifetimes (weeks to months) and which typically release the reactive forms, HOx, NOx and ClOx via photolysis or thermal decomposition. Examples of reservoir molecules include nitric acid ($HNO_3$), nitrogen pentoxide ($N_2O_5$) and hydrochloric acid (HCl).

Tian et al. (2014) suggested that high (FUV/NUV) ratios characteristic of M-dwarf stars (with ratio values up to x100 higher than the Sun) could favor abiotic $O_2$ from $CO_2$ photolysis in planetary atmospheres since firstly, strong FUV favors $CO_2$ photolysis and secondly, weak NUV favors low release rates of HOx and NOx from their reservoirs hence weak regeneration of $CO_2$. To develop these ideas further, it would be interesting to quantify with climate-chemistry modeling studies the effect of varying (FUV/NUV) upon the release of HOx, NOx and ClOx in different background atmospheres and to clarify the role(s) of different reservoir molecules over a range of UV environments considering e.g. weakly-bonded reservoirs such as nitrous acid (HONO) and hypochlorous acid (HOCl) which photolyse in the visible and NUV respectively; medium-strength reservoirs such as chlorine nitrate (ClONO$_2$), peroxynitric acid (HO$_2$NO$_2$) and nitric acid (HNO$_3$) which photolyse in the UV and strongly-bound reservoirs such as HCl and H$_2$O which photolyse in the UV and EUV. On Mars and Venus, HOx and ClOx cycles respectively have been suggested to play an important role in regenerating CO$_2$ (see e.g. Yung and DeMore, 1999 and references therein). Stock et al. (2017) modelled catalytic cycles which operate in the modern Martian atmosphere. Grenfell et al., (2010) suggested an alternative route to converting CO into CO$_2$ via heterogeneous oxidation of CO on the surface of transition metal containing minerals such as hematite on (hot) planetary surfaces.

Could CO-dominated atmospheres build-up in-situ via CO$_2$ photolysis in the gas-phase? This would require firstly, CO$_2$-rich atmospheres which photolyse into CO plus O under conditions where the reverse reaction (reforming CO$_2$) is suppressed due to a lack of HOx, NOx or ClOx (see Selsis et al., 2002). In general, HOx is formed in atmospheres with trace amounts of water vapor in the presence of UV whereas NOx can build up in atmospheres containing N$_2$ with some O$_2$ in the presence of EUV or/and cosmic rays and lightning. Secondly, build-up of a CO-dominated atmosphere would be favored by atmospheres in which the abiotic O$_2$ formed as a by-product of CO$_2$ photolysis could be removed by weathering (see 5.3 and Zahnle et al., 2008).

In summary, CO atmospheres could therefore be favored for atmospheres with the following conditions: (1) strong early CO$_2$ outgassing, (2) dry atmospheres (suggesting low HOx), (3) weak dynamical transport cells (suggesting weak lightning), (4) protected from cosmic rays. Attaining a CO-dominated atmosphere in this way would require removal of the co-produced abiotic O$_2$ via e.g. weathering. More work is required however to explore this parameter range. Zahnle et al. (2008) (their Figure 11) explored a parameter range where CO is favored depending upon humidity and temperature. They also noted that CO atmospheres on early Mars, if they existed, could have left behind metal carbonyls in the Martian soil, which would be a proxy for their existence.

**5.5.2 Outgassing**

Outgassing of CO directly could also favor formation of CO-dominated atmospheres on early terrestrial exoplanets if mantle fugacity is suitable (Miyakawa et al., 2002). On modern Earth, even if carbon were outgassed as CO, it would nevertheless be rapidly removed due to its rather short lifetime of ~2 months (Novelli et al., 1994) in Earth's damp troposphere which proceeds mainly (~85% Khalil and Rasmussen, 1990) via HOx-catalysed CO-oxidation into CO$_2$ with smaller contributions from surface losses and diffusion into the stratosphere.

**5.5.3 Thermodynamics**

Observed values of CO, CH$_4$ and H$_2$O in the atmospheres of Hot Jupiters (e.g. Tinneti et al., 2007; de Kok et al.,2013) and in particular the [CO/CH$_4$] ratio have presented a challenge for photochemical-climate-transport models to interpret (see e.g. Venot et al., 2014 and references therein). More work for example is required to understand better the potential role of e.g. vertical transport, metallicity and photochemistry upon [CO/CH$_4$] in such atmospheres. CO is generally the thermodynamically favored form of carbon compared to CH$_4$ in (Ultra) Hot Jupiters [U(HJs)] for pressure regions typically sampled during emission (Kataria, et al., 2016) such that [CO/CH$_4$] ~10 for cooler HJs increasing to up 10$^{-8}$ for UHJs and where the sum [CO+CH$_4$] constitutes ~1% of the atmosphere. More work is required however to understand better the potential role of e.g. vertical transport and photochemistry upon the [CO/CH$_4$] ratio.

Regarding warm and hot SEs, modeling work by Hu et al. (2014) (their Figures 5 and 6) suggested that chemical equilibrium favors CO-dominated atmospheres e.g. for GJ1214b and for 55 Cancri e in the pressure region (1-100mb) (assuming 0.5<X$_{[C/O]}$<2 and X$_{[H]}$<0.5, where 'X' denotes mole fraction) generally observable by transmission spectroscopy. More work is required to explore the role of photochemistry and the importance of vertical transport which can upwell CO from lower hotter regions where it is thermodynamically favored.

**5.5.4 Other Processes**

Additional sources of atmospheric CO potentially important on e.g. Early Earth include delivery of CO directly via comets (e.g. Lupu et al., 2007). Another suggested source of CO occurs on the surface of hot metallic iron from impact plumes which can reduce CO$_2$ into CO (Kasting, 1990). Finally, it is interesting to note that Miyakawa et al. (2002) suggested there is no geological proxy to *preclude* CO-dominated atmospheres on Early Earth and that such atmospheres could even have favored prebiotic synthesis.

**5.6 Sulphur-containing atmospheres**

Sulfur and its compounds have fascinatingly diverse applications in planetary science. This diversity stems from sulfur's ability to adopt a range of redox and valence states which enable the formation of stable chemical compounds applicable to a wide range of chemical and biological systems extending from the planetary interior to the lithosphere, hydrosphere and atmosphere.

**5.6.1 Solar System**

Regarding Earth, the modern and ancient global sulfur cycles were reviewed by e.g. Brimblecombe (2014) and Fike et al. (2015) respectively. In Earth's core, seismology measurements suggest that light alloying elements including sulfur could be present (see e.g. review by Poirier, 1994; sulfur compounds were also suggested to occur in the core of other planets such as Mercury (Rivoldini et al., 2009) and Mars (Schubert and Spohn, 1990). In Earth's upper mantle, sulfur compounds are estimated to make up 300-400ppm (von Gehlen, 1992). On Earth's surface sulfur is cycled from lithosphere to interior via burial of oxidized (sulfate) and reduced (sulfide) and returned to the atmosphere via outgassing of e.g. $H_2S$ and $SO_2$. Analyzing deviations from standard mass dependent fractionation of isotope ratios in sulfur sediments from the Early Earth (see e.g. Harman et al., 2018) can be used as a proxy for atmospheric oxygen abundance.

In modern Earth's atmosphere, important surfaces sources of sulfur-containing species include: $SO_2$ (60-80 Tg S/yr) and $H_2S$ (7-8 Tg S/yr) arising mainly via anthropogenic and volcanic emissions and dimethyl sulfide (DMS) ($CH_3$-S-$CH_3$) (13-25 Tg S/yr) arising mainly via biogenic emissions (Sheng et al., 2015). Potentially complex gas-phase oxidation cycles in the troposphere (see for example von Glasow and Crutzen, 2004) involving oxidants such as hydroxyl (OH) and ozone ($O_3$) then lead to formation of gas-phase sulfur trioxide ($SO_3$) which quickly undergoes the gas-phase reaction: $SO_3+H_2O+M \rightarrow H_2SO_4+M$ (where 'M' denotes any species needed to carry away excess vibrational energy). The gas-phase products sulfuric acid quickly reacts with liquid water to form sulfate aerosol. The so-called Junge layer (Junge et al., 1961) denotes Earth's stable, globally-distributed sulfate aerosol peaking around 18-22km which influences climate, clouds and composition. Faloona (2009) review uncertainties in the processes affecting atmospheric sulfur in the Earth's marine boundary layer.

On Venus, a thick cloud layer extends from ~48 to 70km which likely consists mostly of sulfate aerosol droplets formed from $SO_2$ oxidation. These clouds strongly influence atmospheric albedo, climate and composition. Below the clouds at 35-40km, VEx data (Svedhem et al., 2007) suggested 50-500ppmv $SO_2$ and 2-8 ppmv COS. Above the clouds, Mahieux et al. (2015) suggested significant variability of $SO_2$ (from ~tens to hundreds of ppbv) based on Venus Express (VEx) data. Sulfur-based catalytic gas-phase cycles have been proposed above the clouds (Yung and Demore, 1999, their chapter 8) to drive recycling of CO back into $CO_2$ on Venus. Recent observations of the glory phenomenon by Vex (Markiewicz et al., 2014) suggested that elemental polysulfur chains or/and iron sulfide could be the unknown UV absorber.

On Mars, the presence of mineral sulfates (see e.g. review by King and McLennan, 2010), (which were also suggested by the Curiosity Rover, Nachon et al., 2017), together with a general lack of carbonates, suggest that sulfur cycles could have played an important role in the lithosphere-atmosphere system on early Mars. Some works e.g. Tian et al., (2009), Kerber et al., (2015) modelled the influence of atmospheric $SO_2$ and sulfate aerosol upon climate on early Mars. Results suggested e.g. that sulfate aerosol cooling could prevent $SO_2$ greenhouse heating from producing habitable surface conditions. Halevy and Head (2014) used an updated model to reinvestigate this problem. They found that on a cold, early, dusty Mars the net effect of $SO_2$ outgassing could be warming yet could be cooling if the atmosphere is clear, consistent with the findings in Tian et al. (2009).

In the outer Solar System, observations of sulfur-containing atmospheric species are rather sparse. Wong et al. (2004) suggested (68-110) ppmv of $H_2S$ in the deep Jovian atmosphere (from 8.9-11.7 bar) based on a revised Galileo Probe analysis. Sulfur-containing species could make up an important constituent of clouds e.g. $NH_4SH$ detected on Saturn (Baines et al., 2009) and $H_2S$ on Uranus (Irwin et al., 2018). Regarding the icy moons, $SO_2$ has been detected on Io as discussed above (see Table 1). Moses et al. (2002) modelled photochemical responses of sulfur species during volcanic eruptions on Io. Some studies hypothesized biogeochemical cycles including sulfate occurring in the deep oceans of icy moons such as Europa (Zolotov and Shock, 2004).

Regarding exoplanets, what range of sulfur-containing atmospheres could be feasible? On terrestrial planets under oxidizing conditions (e.g. on Earth or above the cloud-top on Venus) in the presence of UV, reduced sulfur compounds could form oxidized species such as $SO_3$. This species quickly reacts with trace amounts of water vapor to form sulfate aerosol. Under reducing conditions (e.g. on giant planets), on the other hand, reduced sulfur species such as $NH_4SH$ or/and elemental sulfur could condense to form aerosol. The model study by Hu et al. (2013) for example, who investigated $H_2$-dominated atmospheres for SEs for example suggested the presence of sulfur which can form multiple structures including a stable $S_8$ crown (whose spatial arrangement leads to low strain on the S-S bonding) which could condense and form aerosol. Kaltenegger and Sasselov (2010) discuss spectral signals for sulfur species assuming hypothetical exoplanetary geochemical sulfur cycles. Regarding gas planets, Zahnle et al., (2016) also noted the potential importance of $S_8$ hazes in their model study of the warm Jupiter 51 Eri b. Zahnle et al. (2009) discussed sulfur photochemistry on hot Jupiters (see also Gao et al., 2015). Loftus et al. (2019) evaluated the co-existence of a thick (observable) sulphur aerosol haze layer together with a significant ocean on a rocky exoplanet, including the fundamentals of the sulphur cycle. Results suggest that such an observable haze layer is not compatible with such an ocean.

Could sulfur containing or even sulfur-dominated atmospheres be common in nature? There are theoretical arguments which suggest that the build-up sulfur-dominated atmospheres could be challenging: in oxidizing atmospheres, sulfur species would be converted to sulfate aerosol which would eventually undergo removal via deposition or washout; in reducing atmospheres e.g. elemental sulfur such as stable $S_8$ could similarly build-up, condense and undergo removal. The extent by which sulfur species can build up depends therefore on the UV intensity and the abundance of water-vapor. Increased sulfur abundances are expected to be favored for weak UV, dry conditions.

**5.7 Nitrogen Atmospheres**

Nitrogen species play a major role in planetary atmospheres, influencing climate, composition and (on Earth) biogeochemical cycling. Modern VEM have surface partial pressures of $N_2$ equal to (3.3, 0.78 and $1.7 \times 10^{-4}$) bar respectively (Yung and DeMore, 1999). VEM interiors are estimated to have (2.2, 3.4, 0.3) (in units of $10^{19}$) kg N respectively (Johnson and Goldblatt, 2015) where values are uncertain by a factor of two to three. The evolution of atmospheric nitrogen on VEM and terrestrial exoplanets was discussed e.g. in Lammer et al., (2019). Wordsworth (2016) suggested that enhanced nitrogen in the Venusian atmosphere compared with Earth could be related to stronger $N_2$ outgassing on Venus, related to enhanced oxidation of the mantle due to enhanced $H_2O$ photolysis followed by H-escape. Goldblatt et al. (2009) and Wordsworth and Pierrehumbert (2013) suggested that atmospheric molecular nitrogen could have helped warm the early Earth to achieve and maintain habitable conditions. A difficulty of this hypothesis is that the atmospheric pressure on Archean Earth might have been too low to support a strong $N_2$-$H_2$ greenhouse effect based on measured raindrop distributions ~3 Ga (Som et al., 2012; Som et al., 2016). The model study of von Paris et al. (2013) suggested that adding 0.5 bar $N_2$ to the early Martian atmosphere led to +13K surface warming. In the present work, we summarize important general processes affecting atmospheric nitrogen evolution with the aim of discussing what possible range of atmospheric $N_2$ abundances could be expected on terrestrial exoplanets. We do not focus here on the so-called waterworlds since potential nitrogen cycles on such worlds are likely very different (see discussion in Lammer et al., 2019).

Important processes affecting planetary nitrogen evolution include (1) *delivery* during planet formation via nitrogen-containing ices present on bodies from the outer solar system (see section 3.4.4). Recent estimates of modern Earth's nitrogen budget suggest ~x50 times the present day Atmospheric Mass $N_2$ (AMN) which is mainly stored in the mantle (~x7 AMN in reduced forms, NHx) and possibly a large reservoir in the core (~x43 AMN alloyed with iron in the form $Fe_xN_y$ as suggested by Laneuville et al., 2018, their Table 1) (see also Marty, 1995); (2) regarding early hydrodynamic *escape* of nitrogen, Lammer et al., (2018) discussed weak loss on early Venus and Earth but efficient loss on early Mars based on nitrogen isotope data; (3) *outgassing* which depends on mantle redox and pH which determine whether nitrogen exists in the molecular state (easily outgassed, favored at pH ~1-3 at ~positive Quartz-Fayalite-Magnetite buffer values in mantle wedges) or in the ammonic ($NH_4^+$) state (easily retained, favored at negative pH in the upper mantle, see e.g. Mikhail et al., (2017) and Zerkle and Mikhail (2017) their Figure 2). Lammer et al., (2019) suggested ambient $N_2$ reached up to ~0.2bar during the Archaean associated with e.g. outgassing at the magma ocean phase; (4) efficient *dissolution* of outgassed $N_2$ into the mantle during the MO phase (Wordsworth, 2016); (5) efficient biological *fixation* and *denitrification* with a rate of ~1-2 $\times 10^2$ Tg N year on modern Earth (Laneuville et al., 2018; see also Canfield, 2010); (6) *lightning* which fixates molecular nitrogen at a rate of ~(3-10)$\times 10^9$ kg N/yr on modern Earth (Laneuville et al., 2018). The resulting nitrogen-containing reactive species form nitrogen oxides or/and nitric acid ($HNO_3$) which is efficiently dissolved by scavenging rain droplets and washed out into the ocean on timescales of weeks to months to form thermodynamically stable nitrate ($NO_3^-$) anions; (7) *cosmic rays* (see discussion below) which can remove $N_2$ and $O_2$ on Earth especially at higher latitudes during strong solar activity (see e.g. Jackman et al., 2005); molecular nitrogen can be returned to the atmosphere via outgassing with an estimated rate (including the contribution from mid-ocean ridges) of up to ~$0.4 \times 10^9$ kg N/yr (Laneuville et al., 2018, their section 2.6); (8) meteoritic impacts can thermolyse atmospheric nitrogen where reaction products form nitrogen oxides and are removed from the atmosphere (Grady et al., 1998); (9) *escape*, depending upon upper atmosphere temperature and incoming XUV on the Hadean Earth which are not well constrained (see discussion in e.g. Lammer et al., 2019).

Regarding theories and proxy data for nitrogen on early Earth, surface partial pressure of nitrogen was likely low during Earth's magma ocean phase due to efficient dissolution of nitrogen from the atmosphere into the hot mantle (see Lammer et al., 2019; Wordsworth, 2016). Som et al. (2012) suggested an upper limit of about twice modern levels for Archaean surface air density based on an analysis of fossilized raindrop imprints.

Kavanagh and Goldblatt (2015) revised the method and suggested an upper limit of about four times the modern value. An updated study by Som et al. (2016) then revised downwards the Archean upper limit for atmospheric surface density, to a value less than half the modern amount. Regarding model studies, Rimmer and Rugheimer (2019) suggested that ~0.3 bar surface pressure on Early Earth would enable sufficient oxidizing conditions in the upper atmosphere to oxidize incoming Archaean micrometeorites, consistent with proxy data. Johnson and Goldblatt (2018) modelled the biogeochemical nitrogen cycle including deep and shallow ocean parameterizations and linking with phosphorous chemistry in the lithosphere. Their results for surface nitrogen partial pressure over Earth's history varied from about (0.1 to 6.0) times the present day atmospheric nitrogen level, depending upon uncertainties in the amount of nitrogen in the bulk silicate Earth (BSE) as well as the timing and magnitude of key processes such as burial, continental evolution, weathering and biological activity. In order for their evolutionary scenarios to finish with the correct present day atmospheric nitrogen (PAN) required (4 to 6) times PAN to be present in the BSE. Stüeken et al. (2016) applied an Earth system box model to study the evolution of nitrogen on Earth. Results suggested that depending on uncertainties in nitrogen burial rates (which can strongly fluctuate based on the geological record) the surface partial pressure of nitrogen can vary from about (0.1 to 1.1) times the present day value.

Regarding exoplanets, Lammer et al. (2019) considered the effect upon atmospheric nitrogen evolution for (a) an Earth-like planet without plate tectonics, (b) a planet with anoxic life, and (c) an Earth-like planet where life becomes extinct. Their results suggested that in case (a) the absence of plate tectonics would hinder mixing of oxidized forms of nitrogen into the mantle. This suggests enhanced abundances of nitrogen in its reduced forms (e.g. $NH_4^+$) in the interior, which are more easily retained hence only a weak conversion to $N_2$ and reduced outgassing of nitrogen. In case (b) anoxic fixation in the absence of denitrification led to low atmospheric nitrogen whereas in case (c) the absence of nitrogen recycling associated with life resulted in atmospheric nitrogen removal via lightning and cosmic rays reading to low atmospheric abundances. Stüeken et al. (2016) also applied a biogeochemical box model to scenarios of anoxic planets with and without life. Their results suggested that whether atmospheric nitrogen could recover to present day levels depended on the rate of burial assumed and whether or not continents were assumed to be present (scenarios without continents did not lead to nitrogen recovery). Laneuville et al. (2018) modeled atmospheric nitrogen evolution on an Earth-like planet without life and suggested variations from ~(0.2 to 8) times the present atmospheric level (PAL) depending mainly on uncertainties in the atmospheric fixation rate due to e.g. lightning and cosmic rays, with a weaker variation of ~(0.6 to 2) PAL arising due to uncertainties in mantle mixing and ocean circulation.

Regarding Earth-like planets orbiting in the close-in HZ of active, low mass stars, the atmosphere could be strongly exposed to high energy particles (Grenfell et al., 2012; Tabataba-Vakili et al., 2016; Scheucher et al., 2018; Herbst et al., 2019, see also Airapetian et al., 2018). Stellar proton events (SPEs) on Earth can dissociate molecular nitrogen and lead to $~2 \times 10^{33}$ NO molecules/day (Jackman et al., 2005, their Table 1). Assuming such events are experienced continuously for an Earth-like planet orbiting an active M-dwarf star, would lead to a removal rate of atmospheric nitrogen due to cosmic rays of 0.04 Tg NO/yr. By comparison, lightning fixation on modern Earth (which without life would remove atmospheric nitrogen within a few hundred Myr without regassing; Laneuville et al., 2018) removes atmospheric nitrogen about x100 slower. For superflaring stars however (see e.g. Davenport, 2016) the cosmic ray fluxes could be $10^3$ to $10^5$ times higher than Earth SPEs, suggesting nitrogen removal fluxes up to ~4000 Tg/yr which would remove Earth's atmospheric nitrogen in ~$10^5$ years. In addition to potentially strong nitrogen removal due to cosmic rays, lightning removal could also be potentially strong on, for example, tidally-locked Earth-like exoplanets which could have strong day to night transport cells, although such effects are not well quantified. Planets with high meteoritic bombardment could fix and efficiently remove atmospheric nitrogen.

In summary, what range of $N_2$ atmospheres could we expect in terrestrial exoplanets? Earth system modeling suggests values from below a tenth of a bar up to several bar over Earth's history. Isotope estimates suggest a large range up to ~43 times the present day atmospheric budget in the Earth's core. The extent and rate at which nitrogen is removed to the core during planet formation needs further study. The main uncertainties in atmospheric processes lie in the distribution of nitrogen in the Earth system and the timing and nature of key processes such as biology and subduction. Model feedbacks such as the effect of e.g. atmospheric oxidation via escape and the influence upon the interior are lacking. A detailed intercomparison of current nitrogen cycle models including their reservoirs and fluxes for clearly defined scenarios would be informative. On a positive note, there is some hope that exoplanet science could help by constraining nitrogen abundances from observing spectroscopic collisional pairs (Schwieterman et al., 2015 as mentioned above) which will be invaluable in disentangling the possible evolution scenarios.

**6. Conclusions**

- Key properties such as the metallicity and early rotation rates of interstellar clouds could vary systematically across the Galaxy. The extent by which this variation could imprint itself upon resulting planetary diversity is currently unclear.

- Although tenuous atmospheres could have weak spectral features they could nevertheless feature large "cometary like" tails and transmission windows down to the surface - making them favorable targets in this respect.

- If water delivery to rocky exoplanets depends mainly on impact-induced de-volatization during formation then water-rich terrestrial worlds could be quite common. If however it depends on a "Grand Tack" type analogue then more work is required to investigate how this mechanism operates for different exoplanetary systems.

- Magma oceans with silicate atmospheres, light primordial atmospheres and outgassing of massive steam atmospheres could be universal stages in the evolutionary track of rocky worlds. How long these phases endure however depends on key processes such as instellation, escape and outgassing.

- The relative occurrence of oxidized atmospheres (e.g. $O_2$, $CO_2$, CO, $H_2O$) compared with reduced atmospheres ($CH_4$ and higher hydrocarbons) and other atmospheric compositions (e.g. sulfur and nitrogen-containing species) depends on numerous processes such as the initial disk metallicity and its dynamics, how efficiently the planetary interior stores reduced compounds and also upon long-term escape, outgassing and biogeochemical cycling.

- Sulfur species can form aerosols in both oxidized and reduced atmospheres. Build-up of such aerosols would eventually be limited by sedimentation and surface deposition. Sulfur-rich atmospheres could be favored where conditions are (a) dry, since water is needed to form sulfate from its precursors, or (b) have weak UV, which leads to suppressed formation of aerosol precursors.

- Nitrogen abundances in the atmosphere vary over two orders of magnitude depending upon uncertainties in interior, subduction and biological processes.

**Acknowledgements**
This project has received funding from the European Research Council (ERC) under the European Union's Horizon 2020 research and innovation program (grant agreement No. 679030/WHIPLASH). This project is partly supported by the International Space Science Institute (ISSI) in the framework of an international team entitled "Understanding the Diversity of Planetary Atmospheres." This project has received funding from the European Union's Horizon 2020 research and innovation program under the Marie Sklodowska-Curie Grant Agreement no. 832738/ESCAPE. M.T. acknowledges funding from the Gruber Foundation. This work has been carried out within the framework of the NCCR PlanetS supported by the Swiss National Science Foundation. This research has made use of NASA's Astrophysics Data System. LN and JLG acknowledge support from the Deutsche Forschungsgemeinschaft (DFG, German Research Foundation) Project ID 263649064 TRR 170. This is TRR 170 Publication No. 87. The authors thank ISSI Teams 370 and 464 for fruitful discussions. Mareike Godolt acknowledges support by the German science foundation through project GO 2610/1-1 and the priority program SPP 1992 "Exploring the Diversity of Extrasolar Planets" (GO 2610/2-1).